\documentclass[fleqn,usenatbib]{mnras}

\usepackage{newtxtext,newtxmath}
\usepackage[T1]{fontenc}

\DeclareRobustCommand{\VAN}[3]{#2}
\let\VANthebibliography\thebibliography
\def\thebibliography{\DeclareRobustCommand{\VAN}[3]{##3}\VANthebibliography}

\usepackage{graphicx}	% Including figure files
\usepackage{amsmath}	% Advanced maths commands
\usepackage[justification=justified]{caption}
\usepackage{subcaption}
\usepackage{ragged2e}
\usepackage{xspace}
\usepackage{float}

\title[PLATO False Positive Detections]{Population Study of Astrophysical False Positive Detections in the Southern PLATO field}

\author[J. C. Bray et al.]{
J. C. Bray,$^{1,2}$\thanks{E-mail: john.bray@open.ac.uk)}
U. Kolb,$^{1}$
P. Rowden,$^{3}$
Robert Farmer,$^{4}$
A. B$\ddot{\textrm{o}}$rner $^{5}$
O. Kozhura$^{1}$
\\
$^{1}$The Open University, School of Physical Sciences, Walton Hall, Milton Keynes MK7 6AA\\
$^{2}$Department of Physics, University of Auckland, Private Bag 92019, New Zealand\\
$^{3}$Royal Astronomical Society, Burlington House, Piccadilly, London W1J 0BQ\\
$^{4}$Max-Planck-Institut für Astrophysik, Karl-Schwarzschild-Straße 1, 85741 Garching, Germany\\
$^{5}$German Aerospace Centre, Institute of Optical Sensor Systems, Rutherfordstr. 2, 12489 Berlin, Germany
}

\date{Accepted XXX. Received YYY; in original form ZZZ}

\pubyear{2015}

\begin{document}
\label{firstpage}
\pagerange{\pageref{firstpage}--\pageref{lastpage}}
\maketitle

% Abstract of the paper
\begin{abstract}
For the upcoming PLAnetary Transits and Oscillation of stars (PLATO) satellite mission, a large number of target stars are required to yield a statistically significant number of planet transits. Locating the centres of the long duration observational phase (LOP) fields closer to the Galactic plane will increase the target star numbers but also the astrophysical false positives (FPs) from blended eclipsing binary systems. We utilise the Binary Stellar Evolution and Population Synthesis (\textsc{BiSEPS}) code, to create a complete synthetic stellar and planetary population for the proposed southern LOP field (LOPS0), as well as for a representative portion of the northern LOP field (LOPN-sub). For LOPS0 we find an overall low FP rate for planets smaller than Neptunes. The FP rate generally shows little variation with Galactic longitude ($l$), and a modest increase with decreasing Galactic latitude ($|b|$). The location of the LOPS field centre within the current allowed region is not strongly constrained by FPs. Analysis of LOPN-sub suggests a markedly increased number of FPs across the full range of planet radii at low $|b|$ resulting in approximately twice the \%FP rate in the LOPN-sub compared to the corresponding southern field segment in the planet radius range $-0.2<\log(R/$R$_{\oplus})\leqslant0.4$. However, only a few percent of fully eclipsing FPs in LOPS0 in this radius range have periods between 180 and 1,000 days so the vast majority of FPs are expected to be outside the period range of interest for PLATO.
\end{abstract}
\begin{keywords}
methods: numerical -- eclipses -- planets and satellites: detection -- planets and satellites: terrestrial planets -- stars: binaries: eclipsing
\end{keywords}

%%%%%%%%%%%%%%%%%%%%%%%%%%%%%%%%%%%%%%%%%%%%%%%%%%

%%%%%%%%%%%%%%%%% BODY OF PAPER %%%%%%%%%%%%%%%%%%

\section{Introduction}

The European Space Agency (ESA) satellite mission, PLAnetary Transits and Oscillation of stars (PLATO), is currently well into the construction phase with the satellite tentatively scheduled for launch in 2026. While other missions such as Kepler \citep{Batalha:2010aa} and more recently the Transiting Exoplanet Survey Satellite (TESS) \citep{Ricker:2015aa}, have made a significant number of exoplanet discoveries, the PLATO mission is arguably unique in that it has the capability of detecting a large number of Earth-like planets orbiting Sun-type stars \citep{Rauer_2014}. 
Like TESS and Kepler before it, PLATO aims to find exoplanets using the planetary transit method where a reduction in flux from the target star is detected as a planet, or planets, transit across it. Such a detection requires that the planet and the star are aligned when viewed from the satellite. Due to the low probability of this occurring, especially for small planets at large orbital distances such as Earth-like planets around Sun-type stars, a large number of target stars need to be observed if a statistically significant sample of transiting exoplanets is to be detected. As an indication, the 2014 PLATO mission proposal set a target of $>267,000$ stars to be observed for the two long duration observational phases (LOPs) \citep{Rauer_2014}. The added constraints introduced by the resolution of the detectors, the noise to signal ratio required for a successful detection and the need for ground-based radial velocity follow-up measurements in particular, mean that the target stars need to be relatively bright, ideally within the priority magnitude range of $m_v<16$. 
  
   The current proposal for the LOPs is to observe two fields of approximately 2,232 square degrees, one located in the Northern Hemisphere (the LOPN), and the other in the Southern Hemisphere (the LOPS), each for a duration of two years, however the final observing strategy is not expected to be finalised until two years before the launch. The two-year minimum observation period is designed to fulfil one of PLATO's primary goals, which is to find Earth analogue planets \citep{Rauer2018}. 
   
   With such a large field of view (FOV), and short focal length, a large pixel scale is required to cover the field. For PLATO this translates to an area of 15 by 15 arc seconds (15$^{\prime\prime} \times 15^{\prime\prime}$) per pixel. In addition, to reach the large number of target stars required, the observational fields must be located in regions with relatively high stellar densities. Logically, the denser the observational field the higher the likelihood of astrophysical false positives from blended eclipsing binaries, since the foreground and background fields are also proportionately more dense. 
   
   In this work, we use the proposed southern PLATO LOP field (which we refer to as LOPS0), as a case study to quantify the dependency of both planetary transits (PTs) and astrophysical false positives (FPs), on Galactic longitude ($l$), and latitude ($b$). Using these dependencies and the clearly stated required minimum number of target stars, we seek to identify the optimal $l$ and $b$ pointing to minimise the percentage of FP detections (\%FP) which we define as the the number of FPs divided by the number of FPs plus the number of PTs, expressed as a percentage. 
   
   This paper is set out as follows: In Section \ref{tpm} we present a brief description of the PLATO mission relevant to our research. In Section \ref{pst} we describe the stellar evolution and population synthesis tools used to assess the expected PT and FP occurrences. We also examine the planetary and stellar distributions derived from the Kepler field and how these distributions can be used to calibrate our synthetic LOPS0. In Section \ref{spfa} we outline our method of evaluating PTs and FPs for LOPS0 and present our results for this field. In Section \ref{taspf0} we fit functions to the PTs and FPs by $l$ and $b$ to evaluate how the location of the field centre affects these results. In Section \ref{extrap} we use the functions derived in Section \ref{taspf0} to model the \%FP for different $l$ and $b$ field centre locations within the allowed region. In Section \ref{LOPSvsLOPN} we compare representative sub-fields of LOPS and LOPN, and finally in Section \ref{dc} we discuss our findings and present our conclusions.

\section{The PLATO mission}\label{tpm}
Below we outline the details of the PLATO mission relevant to our research. For a more comprehensive account of the mission and its aims refer to \cite{Rauer_2014, Rauer2018} and references therein. While we recognise that other phenomenon, such as stellar activity, can also result in false positive planetary transit signals, in this research we focus only on FPs resulting from eclipsing binaries blended in the same pixels as the target stars.

\subsection{Observing Strategy}
The PLATO mission is a four-year mission comprising two LOPs, one in the Southern Hemisphere and one in the Northern Hemisphere. Each of the LOPs will have a two-year duration. If the mission is extended, it is envisaged that a number of `step-and-stare' phases will be added, covering additional areas of sky but for much shorter time intervals, most likely 2-5 months each. 

\begin{figure*}
\centering
\includegraphics[width=\hsize]{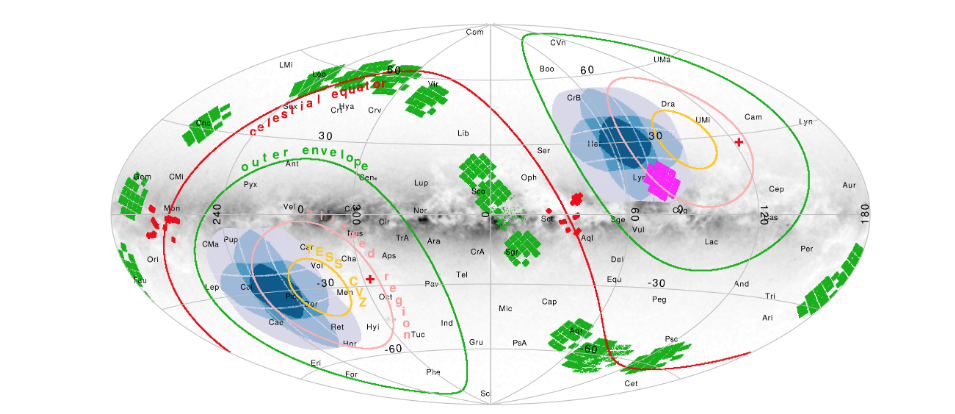}
\caption{Aitoff projection in Galactic coordinates of the proposed Northern and Southern PLATO fields (blue), the area of the allowed PLATO field centres (shown as pink circles) and the total area covered by the allowed PLATO fields (shown as green circles). The CoRoT fields are shown as red squares, Kepler in magenta, K2 in green and the TESS continuous viewing zones as yellow circles.  \citep{Valerio2021}}
\label{fig:PLATOAllowedfields}
\end{figure*}

The proposed LOP fields are 2,232 square degrees with both constrained to have their centres located within spherical caps with the ecliptic coordinate $|\beta| > 63^{\circ}$. While the final field location will not be confirmed until two years before launch, the working assumption within the consortium is that the LOP field centres will not be significantly different from the two originally proposed LOP field centre locations, the Northern Hemisphere LOP (LOPN), centred on Galactic coordinates $l = 65^{\circ}$ and $b = 30^{\circ}$ and the Southern Hemisphere LOP (LOPS), centred on $l = 253^{\circ}$ and $b = -30^{\circ}$. Figure \ref{fig:PLATOAllowedfields} shows the locations of the proposed LOPN and LOPS fields along with the allowed regions for the LOPN and LOPS field centres and the maximum outer bounds. For context the corresponding footprints of the CoRoT, Kepler and K2 surveys, as well as the TESS continuous viewing zones, are also shown.\textbf{}

\subsection{PLATO target stars}
The PLATO consortium has created the PLATO input catalogue (PIC) using data predominantly from Gaia DR2 (RD05) \citep{Montalto2021}. PIC1.1.0 includes PICtarget110, which identifies target stars in both the LOPN and the LOPS. It is envisaged that updated versions of the PIC will be released as more data is released from the Gaia mission and incorporated into the PIC. 

The largest target group P5, is defined as stars with magnitude m$_v\leqslant13.0$ of type F5 to late K with a maximum temperature of 6775K and a minimum temperature of 3875K \citep{PLATORedBook}. For this research we concentrate on the P5 target star group and predominantly on the proposed LOPS which we refer to as LOPS0. This location and target group have been selected as they provide the largest number of stars and hence will produce the most statistically significant results. 

Our analysis focuses on the possible blending of the target stars with background (or foreground), eclipsing binary systems. We consider only  non-grazing eclipses of binary systems which, when blended with flux from the target star mask area, will produce a light curve shape and flux variation similar in magnitude to a planetary transit. 

\section{Stellar evolution and population synthesis; tools and calibration}\label{pst}
To determine the \%FP in LOPS0, we utilise the Binary Stellar Evolution and Population Synthesis (\textsc{BiSEPS}), stellar models to create synthetic single and binary stellar populations for LOPS0 in its entirety. 

To approximate the target star mask area of  \cite{Marchiori:2019aa}, we subdivide LOPS0 into squares of $30^{\prime\prime} \times 30^{\prime\prime}$ or four PLATO pixels. We then identify each four-pixel area where a target star could be found, i.e. each four-pixel square where the m$_v\leqslant 16$. We refer to these as PLATO `super-pixels' and assign each single star in these super-pixels a single planet with a random orbital inclination angle. Where our simulations show that the angle of inclination for any system results in a planet transiting its host star, the flux reduction is calculated. If the flux reduction exceeds the total instrument noise, as calculated in Section \ref{pta}, we class it as a detectable transit event and the corresponding apparent planet radius and orbital period is recorded taking into account blending from the background flux in the super-pixel. Similarly, every binary system in our super-pixel is assigned a random orbital inclination angle and a similar analysis is carried out for all non-grazing stellar eclipses. 

\subsection{Stellar Evolution - The \textsc{BiSEPS} code}
\textsc{BiSEPS} is a population synthesis tool developed by \cite{Willems_2002}. It uses large libraries of different metallicity single and binary stellar models to create Galactic populations. The stellar models are created using simplified binary evolution algorithms based on the prescriptions of \cite{Hurley_2000,Hurley_2002}. The algorithms consider mass loss from stellar winds, angular momentum loss via gravitational waves, magnetic braking and Roche-lobe overflow. The code has been utilised in a number of stellar population studies \citep{Willems_2004, Willems_2006, Davis_2010, PL-Farmer_2013}. Readers wanting more detail on the code should refer to the above studies.

The initial mass parameter space for the primary stars ($M_1$), is divided into 50 evenly spaced logarithmic bins between 0.1M$_{\odot}$ and 20M$_{\odot}$ and into 300 evenly spaced logarithmic orbital separation bins between 3R$_{\odot}$ and $10^6$R$_{\odot}$. The secondary stars ($M_2$), are chosen from the same parameter space as $M_1$ but only systems where $M_1 > M_2$ are evolved. Initial orbits are assumed to be circular and subsequent orbits are kept circularised at each time step. The equivalent circular period can be substituted because, as shown by \cite{Hurley_2002}, orbits generally circularise prior to Roche-lobe overflow so in binaries with the same semi-latus rectum the outcome of the interactions is virtually independent of eccentricity.

For practical reasons, single stars are modelled as binary systems with an orbital period of $10^7$R$_{\odot}$. The primary mass parameter space ($M_1$) is divided into 10,000 evenly spaced logarithmic bins between 0.1M$_{\odot}$ and 20M$_{\odot}$ with the secondary star mass ($M_2$), set at 0.1M$_{\odot}$. 

\subsection{Population Synthesis}
Population synthesis in \textsc{BiSEPS} is carried out by randomly selecting single and binary models from the libraries created above. We assume a binary fraction of 0.5 and assign probabilities for each model using the primary star mass ($M_1$), and an initial mass function (IMF) following \cite{KIMF} with $dN/dM \propto M^\Gamma$ as follows;\\

$\Gamma = 
\left \{
    \begin{array}{l l}
    -1.3  \text{ for} & M_1<0.5\text{M}_{\odot}\\
    \\
    -2.2  \text{ for} & 0.5\leqslant M_1<1.0\text{M}_{\odot}\\ 
    \\
    -2.7 \text{ for} & 1.0\text{M}_{\odot}\leqslant M_1\\        
    \end{array}
\right.$\\

A Galactic structure is created assuming a thin disc with a metallicity of ${Z}=0.02$ \citep{Haywood_2001}, embedded in a thick disc with a metallicity of ${Z}=0.0033$ \citep{Gilmore_1995}. The stellar density of both discs is modelled assuming the double exponential distribution outlined below;
\begin{equation}
\Omega(R_g,z)=\frac{1}{4\pi h^{2}_{r}h_{z}} exp\left(\frac{-R_g}{h_{R}}\right)exp\left(\frac{-|z|}{h_{z}}\right)
\end{equation}
with $h_{R}$ = 2.8kpc and $h_{z}$ = 300pc for the inner disc and $h_{R}$ = 3.7kpc and $h_{z}$ = 1kpc for the outer disc. The Sun is located at $R_g = 8.5$kpc \citep{Reid} and and $z = $30pc \citep{Chen_2001}. 
Star formation is assumed to occur in the thick disc for the first 3Gyr and in the thin disc from 3Gyr to 15Gyr.  Our population includes single and binary stars down to a visual magnitude $m_v=26$.

\subsection{Stellar population and eclipsing binary calibration}
We used the calibrated \textsc{BiSEPS} models created by \cite{PL-Farmer_2013} for their study of the Kepler field. Applying the methods of the Kepler Stellar Classification program \citep{Brown_2011} to their synthetic Kepler field, \cite{PL-Farmer_2013} demonstrated a satisfactory match in $\log(g)$ vs $\log(T_{eff})$ space with the real Kepler input catalogue (KIC),
\citep{Koch:aa}. They then subjected the synthetic KIC to the Kepler target selection process to generate a synthetic Kepler target list. It is on the basis of this synthetic target list that we calibrate the eclipsing binary population of our synthetic population model. 

For periods larger than about 10 days, the synthetic orbital period distribution of eclipsing binaries detectable by Kepler increases with decreasing period and satisfactorily matches the distribution of Kepler Eclipsing Binary Catalog KEBC; \citep{Pr_a_2011, Slawson_2011, Conroy_2014a, Conroy_2014b, lacourse2015kepler, Abdul_Masih_2016, Kirk_2016}. Below this value the KEBC period distribution flattens off while the synthetic distribution continues to rise, resulting in a model over-prediction by about a factor of two for the number of these short-period binaries. The discrepancy is somewhat more pronounced when only those KEBC binaries with a morphology parameter \citep{Matijevi__2012}, less than 0.7 are used, which restricts the sample to systems that are likely detached or semi-detached, i.e. systems which cannot easily be identified as false positives in planet transit searches when blended with other stars. The orbital period discrepancy is related to an under-representation of unequal mass ratio ($q$) short-period binaries relative to the model. 

We remove the discrepancy by introducing a calibration weighting for short-period systems in the synthetic sample ($P<10$ days) which is proportional to $q^{2.5} \times f(\log P)$ where $f(\log P)$ increases effectively linearly from 0.4 at $P=1$ day, to 1 at $P=10$ days (see \cite{PRThesis} for more details). 

\begin{figure}
   \centering
   \includegraphics[width=\hsize]{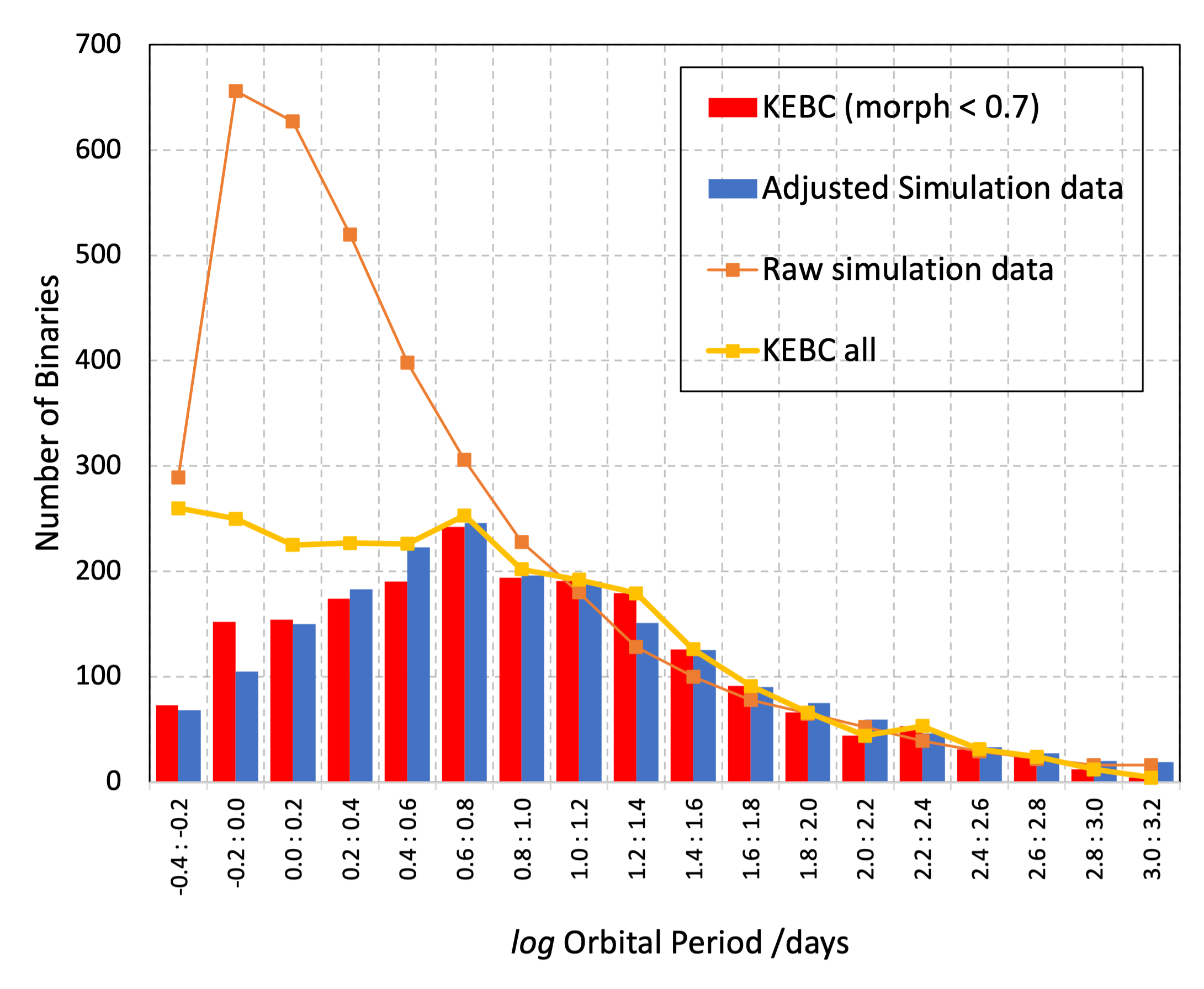}
      \caption{Observed versus simulated Kepler eclipsing binary distributions. The yellow line shows all Kepler eclipsing binaries, the orange line shows the raw simulation data, the red bars show the KEBC binary distribution with morphology parameter $<0.7$ and the blue bars shown the adjusted \textsc{BiSEPS} population synthesis binary distribution.}
         \label{fig:NewPL}
   \end{figure}

Since our analysis was carried out some entries in the KEBC have been updated so we re-visited the KEBC to ascertain if our modified period distribution still provides a good approximation to the KEBC. We reanalysed the data in the KEBC Third Revision (last updated Aug 8, 2019). Figure \ref{fig:NewPL} shows the updated results for the entire KEBC (yellow line) and those with a morphology parameter $<0.7$ (red bars). We have reproduced, on the same plot, our original eclipsing binary period distribution (orange line) as well as the adjusted dataset we described above (blue bars). 

We find our original adjusted eclipsing binary period distribution is still a good approximation for the KEBC, especially the KEBC data with a morphology parameter of $<0.7$.

\subsection{The intrinsic exoplanet distribution}\label{ied}
\begin{figure}
   \centering
   \includegraphics[width=\hsize]{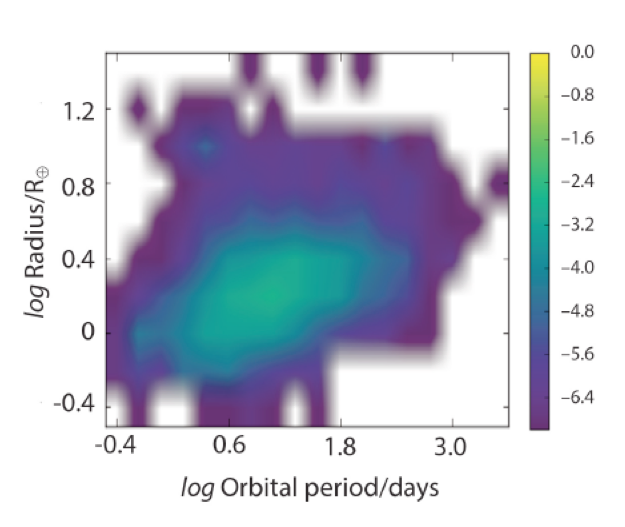}
      \caption{Observed Kepler exoplanet distribution. The contour scale is normalised and displayed in log space}
         \label{fig:PLATOPLObs}
   \end{figure}

We conducted a similar calibration exercise for the intrinsic planet content of the \textsc{BiSEPS} synthetic population model. 
In the synthetic Kepler field, we seeded each single star with one exoplanet, randomly chosen from an initially flat distribution in planet period ($P$) and planet radius ($R$). Each system was assigned a random orbital inclination angle. We then extracted those systems that displayed detectable transits according to the Kepler detector characteristics. We considered both the single `long cadence’ and `benchmark’ estimate sensitivities of the Kepler detectors as outlined in the Kepler instrument manual \citep{KIH}. The extracted detectable synthetic exoplanet distribution over $P$ and $R$ obtained above was then compared against the observed exoplanet distribution shown in Figure \ref{fig:PLATOPLObs} taken from the NASA Exoplanet Science Institute (NExScI)\footnote{ https://exoplanetarchive.ipac.caltech.edu/docs/data.html} catalogue of confirmed Kepler planets. 
We highlight that the relative distribution is over the planet radius of confirmed planets only. The combined group of confirmed and candidate planets is broadly similar for $\log R/$R$_\oplus > 0$, suggesting that completeness may be similar for both samples. The Kepler planet sample is least complete at long periods and small radii, making estimates in this parameter space particularly vulnerable to systematic uncertainties.  Our model results for $\log R/$R$_\oplus<0$ should be seen in this context. 

\begin{figure}
   \centering
   \includegraphics[width=\hsize]{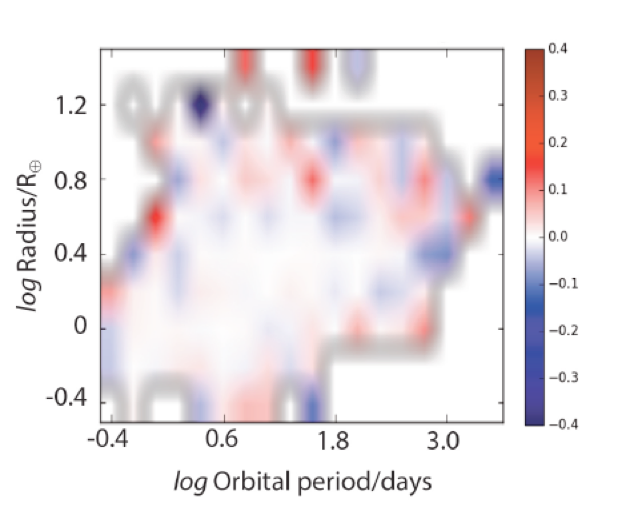}
      \caption{Relative difference between the observed Kepler exoplanet distribution and the detectable synthetic distribution. The contour scale represents the relative difference displayed in $log$ space.}
         \label{fig:PLATOPLObsvsSim}
   \end{figure}

The seeding probabilities were then iteratively adjusted for each $P$ and $R$ bin until a satisfactory match between the synthetic and observed exoplanet $P$ and $R$ distributions were achieved. After applying our selected adjustment matrix, we find the relative difference between the observed confirmed exoplanets and those produced by our synthetic population to be no more than 0.10 dex in the period / radius range of interest to the PLATO mission, namely planet period $\log P>0$ and planet radii $0.0<\log(R/$R$_\oplus)\leqslant0.4$. The comparison is shown in Figure \ref{fig:PLATOPLObsvsSim}. 

With the seeding assumption of one planet per star this procedure leads to a synthetic planet sample which underestimates the total planet count in the Kepler field. While data from microlensing surveys suggests that there is at least one planet per star \citep{Cassan_2012}, more recent studies of planets in the range $0.0<\log(R/$R$_\oplus)\leqslant1.0$ around M dwarfs suggest this number is more likely to be between 1.3 and 2.5 \citep{Sabotta_2021, Feliz_2021}. From the 2,700 or so confirmed planets in the Kepler field nearly one half are in multiple planetary systems. We achieve a match between the observed and simulated planet samples – both in terms of total planet number and in relative distribution over planet radius and planet orbital period - by applying an overall scaling factor of 1.5 to the synthetic sample after the $P$ and $R$ adjustment described above. Again readers wanting more detail are directed to \cite{PRThesis}.

\section{Analysis of the LOPS}\label{spfa}
\subsection{The proposed LOPS (LOPS0)}\label{spfaS1}
\begin{figure}
   \centering
   \includegraphics[width=\hsize]{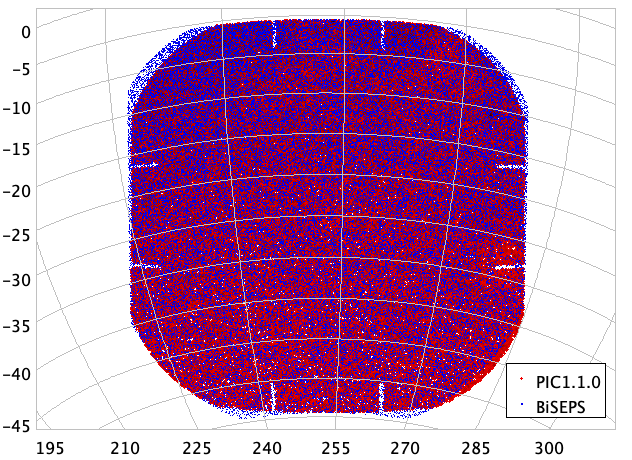}
      \caption{P5 target stars in the proposed Southern PLATO field as defined in the PLATO Input Catalogue 1.1.0 (red), overlaid with the locations of the P5 target stars identified in the synthetic southern PLATO field (LOPS0), used in this research (blue). Galactic longitude x-axis, Galactic latitude y-axis.}
         \label{fig:PLATOfields}
   \end{figure}

\begin{figure}
   \centering
   \includegraphics[width=\hsize]{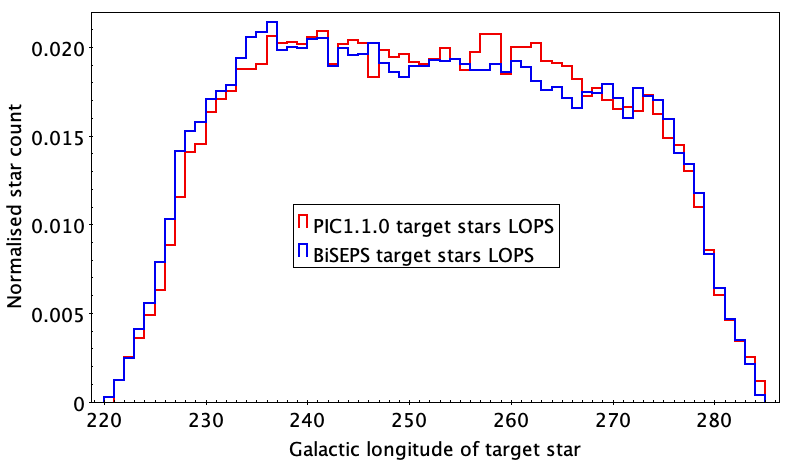}
      \caption{P5 target star distribution with Galactic longitude in the proposed Southern PLATO field as defined in the PLATO Input Catalogue 1.1.0 (red), and in the synthetic southern PLATO field LOPS0 (blue).}
         \label{fig:PLATOfieldslong}
   \end{figure}

\begin{figure}
   \centering
   \includegraphics[width=\hsize]{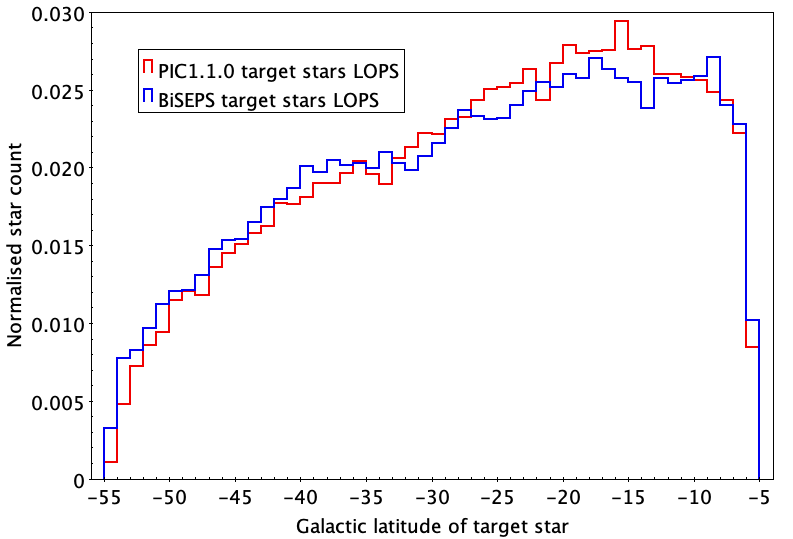}
      \caption{As Figure~\ref{fig:PLATOfieldslong}, but for Galactic latitude.}
         \label{fig:PLATOfieldslat}
   \end{figure}

The LOP fields are not simple geometric shapes so for simplicity of analysis, we have used Galactic coordinates to simulate a square field on the celestial sphere and then shaped the field using an elliptical approximation for each of the four detector groups. While not exact, the resulting synthetic LOPS0 shape and camera overlap pattern centred on $l=253^{\circ}$ and $b=-30^{\circ}$ provides a very good approximation to the LOPS area analysed in PIC1.1.0. A comparison of the LOPS0 area in PICtarget110 and the synthetic LOPS0 area created by this technique is shown in Figure~\ref{fig:PLATOfields} with the normalised target star numbers by Galactic longitude and latitude shown in Figures \ref{fig:PLATOfieldslong} and \ref{fig:PLATOfieldslat} respectively.

The synthetic LOPS0 is created by combining simulations of sub-fields of one degree by one degree in $l$ and $b$. For simplicity we use lines of constant Galactic latitude ($b=-5.5^{\circ}$ and $b=-53.5^{\circ}$), as the upper and lower bounds of our synthetic LOPS0, however, since the width of one degree in $l$ reduces with increasing $|b|$, the number of sub-fields analysed at higher $|b|$ was increased according to the haversine formula. 

The simulation is CPU-time intensive and to make such a task manageable, we have utilised the parallel computing power of the Open University's \textsc{CentOS7} computer cluster to systematically step through the large number of sub-fields. 

To verify that our synthetic LOPS0 target population is representative of the observed population, we carried out a comparison of the number of target stars in the synthetic LOPS0 to that in PICtarget110. 
After removing objects in PICtarget110 with no $T_{eff}$ values, we obtain a total of 140,105 objects with magnitudes $3\leqslant G_v\leqslant13$ and temperatures of $4065$K$\leqslant T_{eff}\leqslant6848$K. Our single star target population for the same magnitude and $T_{eff}$ range contains 76,026 objects. The analysis of the Gaia DR2 catalogue by \cite{Arenou:2018aa} suggests that within $1^{\prime\prime}$ of an object the catalogue is only $\sim$75\% complete and does not exceed $\sim$90\% until separations of greater than $1.5^{\prime\prime}$ are reached. Hence there is a high likelihood that PICtarget110 contains a number of objects, observed as single stars, that our analysis would identify as binary star systems. If we make the first order assumption that any binary system in our synthetic LOPS0 with an angular separation of $1^{\prime\prime}$ or less will not be resolved by Gaia this increases our target star list by 64,230 to 140,256 objects which is in very good agreement with the number in PICtarget110.

As can be seen in Figures \ref{fig:PLATOfieldslong} and \ref{fig:PLATOfieldslat} both the longitude and latitude distributions for the P5 target stars are well reproduced by the BiSEPS population synthesis.

For the first step of our synthetic LOPS0 analysis, we examine each  $30^{\prime\prime} \times 30^{\prime\prime}$ PLATO super-pixel as outlined in Section \ref{pst}. Examining only these super-pixels reduces the analysis area by approximately 50\% at low Galactic latitudes ($b=-6^{\circ}$) and by approximately 90\% at mid-Galactic latitudes ($b=-45^{\circ}$).

\begin{table}
    \centering
            \caption{Variables used in total noise calculation in Section \ref{pta} Key: ppm = parts per million, e- = electrons, ADU = analogue to digital unit, s = second.
            The values for BGN, exposure time and Inverse-gain were obtained from proprietary PLATO mission documentation while readout and jitter noise were obtained from private communications with members of the PLATO Instrument Signal and Noise Budget team}
	        \begin{tabular}{l c c r}
		    \hline\hline
		    Item & ID & value & unit \\
		    \hline\hline
		    Background noise & $BGN$ & 100 & e- / pixel / s\\
		    Readout noise & $RON$ & 57.7 & e-\\
		    Inverse-gain & $G$ & 25 & e- / ADU\\
		    jitter noise& $JN$ & 9 & ppm\\
		    Exposure time & $T_{exp}$ & 21 & s\\
		    \hline
	    \end{tabular}
	\label{tab:1}
\end{table}

\begin{table}
    \centering
    \caption{Detector throughput used in total noise calculation \:   Key: nm = nanometres. Estimates of throughputs by wavelengths obtained from private communications with members of the PLATO Instrument Signal and Noise Budget team.}
    \begin{tabular}{c|c}
    \hline\hline
    Wavelength (nm) & Throughput\\
    \hline\hline
    500 &  0.47\\
    550 &  0.53\\
    600 &  0.59\\
    650 &  0.59\\
    700 &  0.59\\
    750 &  0.52\\
    800 &  0.45\\
    850 &  0.34\\
    900 &  0.22\\
    950 &  0.12\\
    1000 &  0.04\\
    \hline
    \end{tabular}
    \label{tab:2}
\end{table}

\subsection{Planet transit analysis}\label{pta}
This analysis is carried out as follows: we seed every single star (i.e. every stellar source consisting of one star only) in the PLATO super-pixels with one planet according to the intrinsic radius and period distribution derived in Section \ref{ied}. In addition to selecting a planet radius and period, each planet is assigned a random orbital inclination angle. If this angle exceeds the critical inclination angle (signifying a transit), the stellar properties, planet radius, orbital period and orbital inclination are added to the planet transit master file. This creates a master list of all planet transits in the synthetic LOPS0 super-pixels. From this master file, we select single stars with a magnitude $m_v\leqslant13$ and with effective temperatures between $3875$K and $6775$K as the PLATO P5 target star group \citep{Rauer2018}.

The final observable apparent transit depths are calculated by blending the synthetic transit depths with the total flux from the PLATO super-pixel containing the target star. We adopt this approach to approximate the binary mask suggested by \cite{Marchiori:2019aa} as the optimal solution for P5 targets. 

To evaluate the detectability of a transit we use a simple representation of the main characteristics of the PLATO detector and merely determine if the transit depth is larger than the expected noise. A more sophisticated appraisal of the detector and the pipeline processes for extracting transits is beyond the scope of this global study. 

We use the standard CCD equation as a simplified model to calculate the interim instrument noise to signal ratio ($N/S_{int}$), for a four pixel area per camera per exposure as below:

\begin{equation}
N/S_{int}=\frac{\sqrt{F_{\star}+(4\times BGN_{flux}\times T_{exp})+4\times{RON^2}+\left(\frac{G^2}{3}\right)}}{F_{\star}}
\end{equation}

The estimated instrument values used for readout noise $RON$, background noise $BGN$, gain $G$ and exposure time $T_{exp}$ are shown in Table \ref{tab:1}.  The background noise ($BGN$) is approximated as a constant of 100 e- / pixel / s. $BGN$ includes effects such as the contribution from the zodiacal light, and stray light from the Moon, Earth and Sun. Ideally this should be calculated for each super-pixel. A much more in-depth analysis of the detector performance would be required for this which is well beyond the scope of this paper. $BGN$ does not however include unresolved background stars as these are taken into account by virtue of the fact that the synthetic model extends to stars and binaries of 26th magnitude ($m_v=26$), all of which are being added as de-facto contaminants to the target star flux.
$RON$ includes contributions from both the CCD and front end electronics which are added in quadrature. We assume that the point spread function extends over all four pixels of the PLATO super-pixel.
The values for exposure time, inverse-gain and readout noise were obtained from proprietary PLATO mission documentation and private communications with members of the PLATO Instrument Signal and Noise Budget team.

To calculate the total photon count, $F_{\star}$, from the target star received during the exposure, we use the $T_{eff}$ of the target star and the Planck function to calculate its black body spectrum using the throughput by wavelength function shown in Table \ref{tab:2}. We calculate the target star photon count per camera noting the known visual flux for an $m_v=11$ star is 10,856 photons/sec/cm$^2$/nm$^{-1}$ \citep{Johnson}.

The calculated $N/S_{int}$ value is then combined in quadrature with a jitter noise approximation of 9ppm, and then since the cameras are cycling every 25 seconds, the $N/S_{int}$ value for a one-hour integration is reduced further by $\sqrt{25/3600}$.

Our analysis shows that 90\% of target stars have a magnitude $m_v<13$ and 80\% of FPs have a magnitude $m_v>16$. This means that in a `worst case' type scenario where a target star has an $m_v=13$ and a contaminant has a magnitude of $m_v=16$ the effect of adding the background flux into the noise equation would reduce the transit depth by less than 3\%. Further, both the PT and FP analysis would be subject to the same noise which would effectively cancel out leaving the \%FP ratio virtually unchanged. As a result we deem the effect of the background flux negligible and we exclude it from our noise calculation. 

Finally we calculate the number of observing cameras assuming all cameras are functional, so each group contains six cameras. 

\begin{figure}
\centering
\includegraphics[width=\hsize]{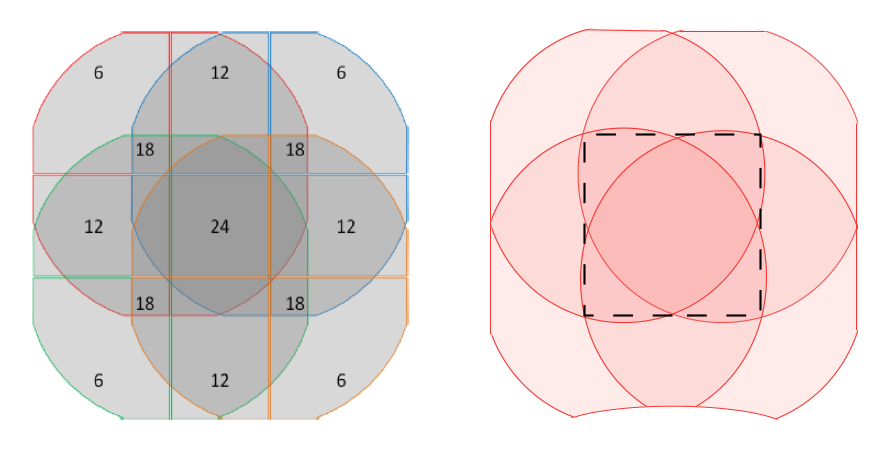}
\caption{The left panel shows the PICtarget110 field of view (FOV), and camera overlap pattern projected onto the celestial sphere (PLATO | Institut d'Astrophysique Spatiale)\textsuperscript{2}. The right panel shows the synthetic FOV and camera overlap pattern calculated in Galactic coordinates, also projected onto the celestial sphere. The boundary of the 18 camera overlap area used in PICtarget110 is shown in the right panel by the dashed black line.}
{\small\textsuperscript{2} https://www.ias.u-psud.fr/en/technical-activities/optics-department/plato}
\label{fig:FOV}
\end{figure}

As shown in Figures \ref{fig:PLATOfields} and \ref{fig:FOV}, the camera coverage and overlap pattern we calculate in Galactic coordinates, provide a good approximation for the FOV used in PICtarget110 with the difference in area being less than 2\%.

For simplicity, the synthetic FOV interior camera group boundaries are approximated as ellipses in Galactic coordinates which closely resemble the observed circular areas when converted to the `on-sky' view. While this results in a slight difference between the area where 18 cameras overlap at the expense of the area where 12 cameras overlap when compared to PIC110 area, the discrepancy is less than 3\%. The boundary of the PIC110 18 camera overlap area is shown by the dashed black line in the right-hand panel of Figure \ref{fig:FOV}. 

We then compare each target star coordinate with each camera group coverage area to calculate the number of camera groups ($N_{cg}$) observing the transit. The $N/S_{int}$ value is then further reduced by $1/\sqrt{N_{cg}\times6}$ to give a final $N/S$ value.

A planetary transit is recorded if the observed transit depth (flux from the star out of transit minus the flux from the star mid-transit), when blended with the total flux from the PLATO super-pixel, exceeds the corresponding detector $N/S$ value calculated above. This is the same threshold we use for false positives resulting from eclipsing binaries. While a one-sigma detection threshold seems optimistic, we find that introducing a higher threshold has little influence on our results (see Section \ref{rspf0} for more detail). 

The detections are binned in a two-dimensional, matrix according to apparent planet radius ($R$), and orbital period ($P$), with this number then multiplied by the normalisation factor of 1.5 obtained in Section \ref{ied}. The seeding process is repeated 15 times and the results averaged. The standard deviation from the 15 simulations is then used as an estimate for the uncertainty for each planet apparent radius/period bin combination. 

\subsection{Eclipsing binary analysis}\label{eba}
For the eclipsing binary analysis, we again interrogate the LOPS0 super-pixels. 
For each target star (regardless of whether there is a planet transit or not, and there may be multiple target stars in a super-pixel), we select an area around the target star of $30^{\prime\prime}$ x $30^{\prime\prime}$ and identify every single star and binary system down to 26th magnitude. We then assign each binary system in this area a random orbital inclination angle and if this angle exceeds the critical orbital inclination angle (signifying an eclipse), the primary and secondary eclipse depths are calculated utilising the JKTEBOP code \citep{Southworth_2004,Southworth_2005, refId0} which includes quadratic limb darkening of \cite{1985A&AS...60..471W}. Finally, we determine the apparent eclipse depth for every eclipse from the reduction of the total super-pixel flux during the respective eclipse, and record the corresponding planet mimic radius.

By using the \textsc{BiSEPS} population files we are able to ensure the 4-pixel mask is universally applied so that even target stars on the very edge of the super-pixel have a full $30^{\prime\prime} \times 30^{\prime\prime}$ area analysed for eclipsing binaries.  

We identify the deepest eclipse depth and if this exceeds the detector noise calculated in Section \ref{pta}, the false positive planet period and mimicked apparent radius is calculated. There may be multiple target stars and multiple background eclipses in each super-pixel and each is analysed separately. Once again, the process is repeated 15 times and the results averaged with the standard deviation from the 15 simulations used as the uncertainty for the FPs in each apparent planet radius/period bin combination. For the purposes of our analysis, we exclude eclipses where only part of the disc of the eclipsing star transits its binary companion. 

Since we only consider  non-grazing planet transits in our detections, we similarly only consider  non-grazing eclipsing background binaries in our false positive analysis. Distinguishing between grazing planetary transits and grazing binary eclipse light curves is not a trivial exercise and while the most sharply “V” shaped transit curves will be determined to be grazing eclipsing binaries, there are very few circumstances where this can reliably be done by visual inspection only.

\subsection{Extinction}
\begin{figure}
   \centering
   \includegraphics[width=\hsize]{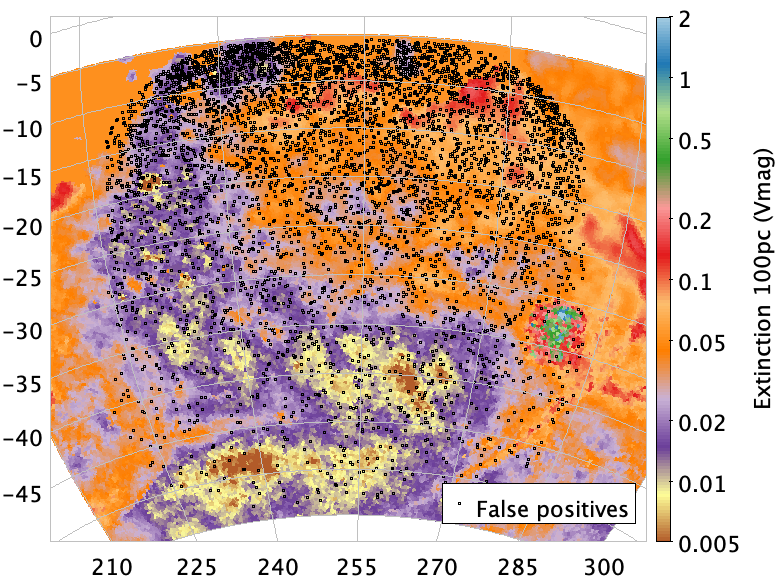}
      \caption{The Drimmel extinction map showing extinction, in $V$-magnitude, at 100pc with false positive locations in the default Southern PLATO field (LOPS0), overlaid in black. Galactic longitude x-axis, Galactic latitude y-axis.}
         \label{fig:EXT100}
   \end{figure}

\begin{figure}
   \centering
   \includegraphics[width=\hsize]{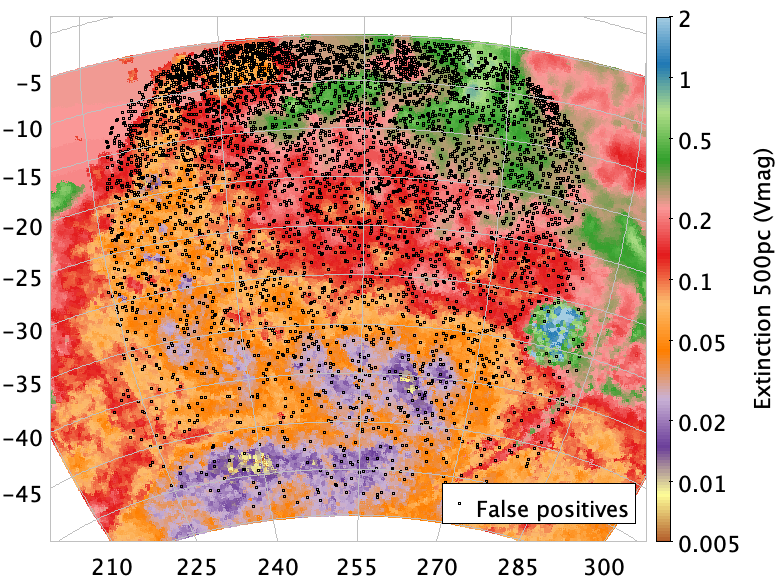}
      \caption{The Drimmel extinction map showing extinction, in $V$-magnitude at 500pc, with false positive locations in the Southern PLATO field (LOPS0), overlaid in black. Galactic longitude x-axis, Galactic latitude y-axis.}
         \label{fig:EXT500}
   \end{figure}

Extinction is calculated following \cite{Drimmel_2003}, who use a 3D dust model scaled from V-band line-of-site extinctions utilizing data from the COBE/DIRBE near infra-red (NIR) instrument. 
The effect of high extinction in the region of Galactic coordinates $l = 280^{\circ}$ and $b = -32^{\circ}$ is reflected in the reduced number of FPs shown in Figures \ref{fig:EXT100} and \ref{fig:EXT500}.

\subsection{Results for LOPS0} \label{rspf0}
\begin{figure}
        \centering
        \includegraphics[width=\hsize]{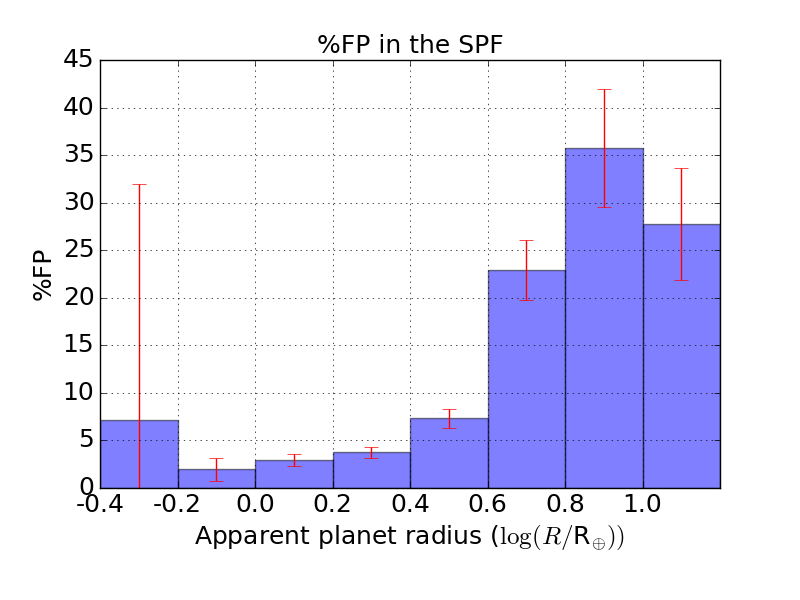}%{FigFPpc.png}
        \caption{The false positive percentage (\%FP), in the default southern PLATO field (LOPS0). Shown by apparent planet radius detected with data from periods in the range $-0.4<\log(P/$days$)\leqslant3.2$ summed and with the \%FP shown as a percentage.}
        \label{fig:PTFPv}
    \end{figure}

The results of our analysis of the synthetic LOPS0 field are shown in Tables \ref{tab:10} and \ref{tab:3} and Figures \ref{fig:PTFPv} and \ref{fig:plots}. For ease of presentation, the period data was combined so the \%FP incorporates all periods from $-0.4<\log(P/$days$)\leqslant3.2$. The \%FPs were calculated on
a simulation by simulation basis to give a set of 15 \%FPs for each apparent planetary radius, these were then averaged and the standard deviation calculated to provide an estimate of the uncertainty in each of the \%FP values. Both PT and FP detections are based on signal greater than noise.

In Figure \ref{fig:PTFPv}, the large \%FP value and error bar for planets with $-0.4 < \log(R/$R$_{\oplus})\leqslant-0.2$ is a result of detections in only four of the 15 simulation runs and in one of those runs there was no planet detections hence the \%FP was 100\%. 

Generally we find a low \%FP in the main region of interest, namely the planets with $-0.2<\log(R/$R$_{\oplus})\leqslant0.4$. Furthermore our research suggests that FPs with periods between 180 and 1,000 days are rare, representing only $\sim$ 2.6\% of the FP numbers. 

The \%FP by planet radius bin (Table \ref{tab:3}), shows that in the $-0.2<\log(R/$R$_{\oplus})\leqslant0.4$ range, the \%FP does not exceed 4\%.

While we show the number of PTs and FPs in our analysis we believe the \%FP ratio is a much more robust statistic to base our results on. To illustrate the effect of a higher detection threshold and the robustness of our \%FP ratio, we show in Table \ref{tab:10} the relative differences in FPs, PT and \%FP using our selected threshold of signal > noise and a second detection threshold where only detections with a signal > five times the noise were recorded. We show the results for the planet radius bins $-0.2<\log(R/$R$_{\odot})<0.8$ and we can see that while the FPs reduce significantly so too do the PTs resulting in a much smaller change, of at most a factor of 2, in the \%FP ratio.

\begin{table}
    \caption{Planetary transits (PT), false positives (FP) and \%FP ((FP/(FP+PT))$\times$100) in the default Southern PLATO field (LOPS0) for planet radii $0.2<\log(R/$R$_{\odot})<0.8$ and for detection thresholds of signal > noise and signal > 5 x noise.}
    \begin{center}
    \begin{tabular}{l c c c c c c c r}
		%\hline\hline
		Planet radius\\ $\log(R/$R$_{\odot})$ & -0.2, 0.0 & 0.0, 0.2& 0.2, 0.4 & 0.4, 0.6 & 0.6, 0.8 & \\
		\hline\hline
		PT (S>N) &  108.0 & 441.2 & 782.8 & 503.8 & 114.1\\
		FP & 2.2 & 13.2 & 29.7 & 39.4 & 33.5\\
		\hline
		\%FP  & 1.9 & 2.9 & 3.7 & 7.3 & 22.9 \\

		\hline\hline
		PT (S>5N)   & 0.7 & 12.5 & 90.4 & 188.4 & 84.7 \\
		FP  & 0.0 & 0.3 & 2.1 & 11.2 & 20.9 \\
		\hline
		\%FP  & -  & 3.9 & 2.2 & 5.7 & 20.0 \\
		\hline\hline
	\end{tabular}
	\label{tab:10}
\end{center}
\end{table}

\section{Trend analysis of the synthetic LOPS0}\label{taspf0}
To identify trends in PTs, FPs and \%FPs, we subdivide the synthetic LOPS0 field into strips which, for ease of reference, we refer to as `horizontal' and `vertical'. For both the horizontal and vertical strip analysis we ignore the detector shape and camera overlap pattern, instead we create rectangular sub-fields of equal area and assign each to be observed by two camera groups.

For the horizontal strips, we subdivide our $48^{\circ}$ Galactic latitude LOPS0 field into 12 strips, each covering $4^{\circ}$ in Galactic latitude. We dynamically set the Galactic longitude boundaries for each latitude setting so that each $1^{\circ}$ latitude sub-strip covers an area of 50 square degrees, giving a total area for each horizontal strip of 200 square degrees. 

For the vertical strips, we work from the centre of the field outwards, setting the inner boundaries of the first two strips at the Galactic longitude $l=253^{\circ}$. We then dynamically set the outer Galactic longitude boundaries for each latitude setting in the strip so that each $1^{\circ}$ latitude strip covers an area of 5 square degrees. 

The process is then repeated with subsequent strips starting where the previous ones ended so that no strip overlaps another, resulting in a total area for each strip of 240 square degrees.

For both the horizontal and vertical strips, we set the lower boundary at the Galactic coordinate $b= -5.5^{\circ}$ and the upper boundary at $b=-53.5^{\circ}$. To remove the confounding effect of increased resolution from the overlapping camera pattern, we arbitrarily assign each sub-field to be observed by two camera groups.

For each vertical and horizontal strip we calculate the PTs, FPs and \%FPs by planet radius bin as described in Sections \ref{pta}, \ref{eba} and \ref{rspf0} respectively. 

We use least squares analysis to calculate the best-fit parameters for linear, exponential and second order polynomial functions and finally select the overall best-fit function according to the Bayesian information criteria (BIC). For our BIC calculations the likelihood is calculated using chi-squared. Uncertainties in our best-fit functions are calculated as the mean of the least-square residuals, $\sqrt{\Sigma (x -\overline{x})^2}/N$. The result of this analysis is shown in Table \ref{tab:4}.

\subsection{LOPS0 - analysis of strips of constant longitude}\label{sclon}
We first discuss the vertical strips which allow us to probe the longitude dependence of the false positive rate across the LOPS0 field. 

As shown in the `V' strips section in the LHS of Table \ref{tab:4}, there is only a single instance where a polynomial fit is preferred over linear or exponential fits and its preference is insignificant.

\onecolumn
\centering
\justify
\begin{figure}
\begin{subfigure}{.48\textwidth}
  \centering
  % include first image
  \includegraphics[width=1\linewidth]{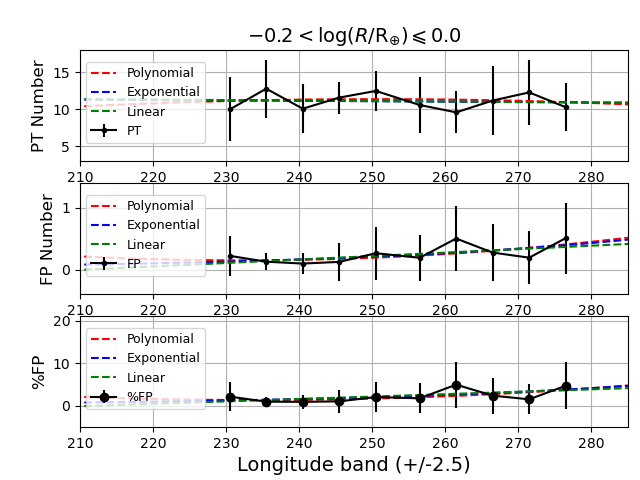}%{M0P2STRNewNoise10.png}  
  \caption{Vertical strip - Planet radii :  $-0.2<\log(R/$R$_\oplus)\leqslant0.0$}
  \label{fig:str0002}
\end{subfigure}
\begin{subfigure}{.48\textwidth}
  \centering
  % include second image
  \includegraphics[width=1\linewidth]{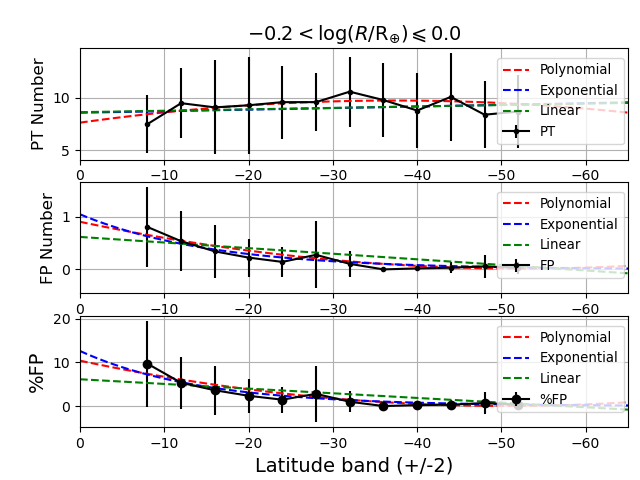}%{M0P2SQRNewNoise10.png}  
  \caption{Horizontal strip - Planet radii : $-0.2<\log(R/$R$_\oplus)\leqslant0.0$}
  \label{fig:sqr0002}
\end{subfigure}

\justify
\begin{subfigure}{.48\textwidth}
  \centering
  % include first image
  \includegraphics[width=1\linewidth]{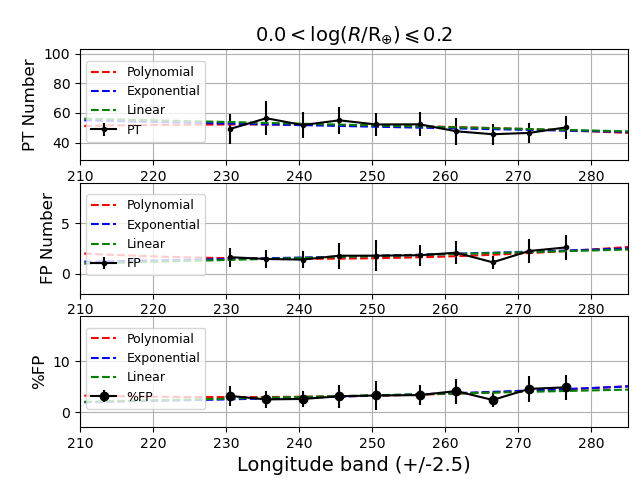}%{0STRNewNoise10.png}  
  \caption{Vertical strip - Planet radii :  $0.0<\log(R/$R$_\oplus)\leqslant0.2$}
  \label{fig:str0204}
\end{subfigure}
\begin{subfigure}{.48\textwidth}
  \centering
  % include second image
  \includegraphics[width=1\linewidth]{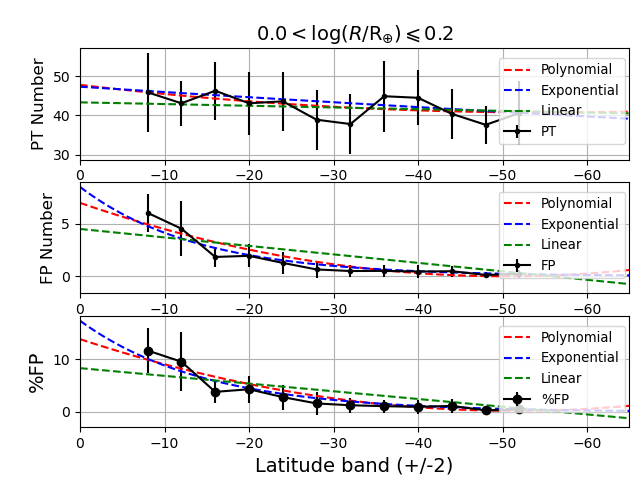}%{0SQRNewNoise10.png}  
  \caption{Horizontal strip - Planet radii : $0.0<\log(R/$R$_\oplus)\leqslant0.2$}
  \label{fig:sqr0204}
\end{subfigure}

%\begin{subfigure}
\justify
\begin{subfigure}{.48\textwidth}
  \centering
  % include first image
  \includegraphics[width=1\linewidth]{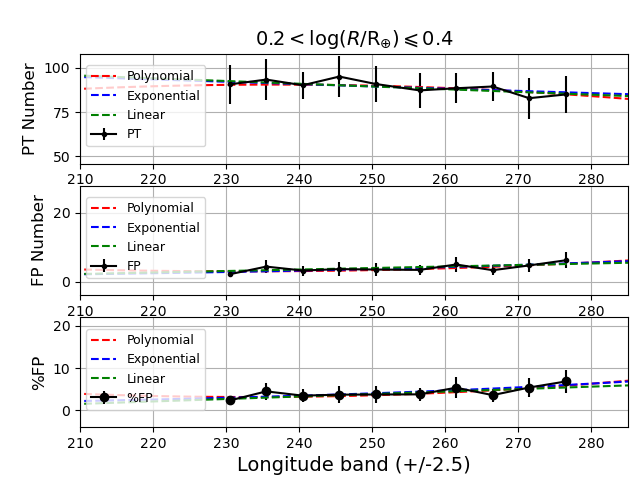}%{0P2STRNewNoise10.png}  
  \caption{Vertical strip - Planet radii :  $0.2<\log(R/$R$_\oplus)\leqslant0.4$}
  \label{fig:str0406}
\end{subfigure}
\begin{subfigure}{.48\textwidth}
  \centering
  % include second image
  \includegraphics[width=1\linewidth]{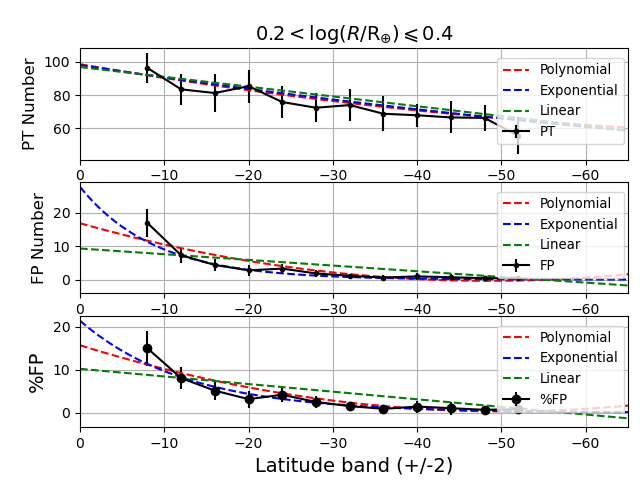}%{0P2SQRNewNoise10.png}  
  \caption{Horizontal strip - Planet radii : $0.2<\log(R/$R$_\oplus)\leqslant0.4$}
  \label{fig:sqr0406}
\end{subfigure}
\caption{Comparison of `vertical' and `horizontal' strip analysis for the three strips covering the radius range $-0.2<\log(R/$R$_\oplus)\leqslant0.4$. The number of planetary transits (PT), top panel, number of false positives (FP), middle panel and the \%FP ((FP/(FP+PT))$\times$100) bottom panel. Plots (a), (c) and (e) show analysis of `vertical' strips, plots (b), (d), (f) show analysis of `horizontal' strips.}
\label{fig:plots}
\end{figure}

\begin{table}
    \caption{Planetary transits (PT), false positives (FP)\\ and \%FP ((FP/(FP+PT))$\times$100) in the default Southern PLATO field (LOPS0)}
    \begin{center}
    \begin{tabular}{l c c c c c c c r}
		%\hline\hline
		Planet radius - $\log(R/$R$_{\odot})$ & -0.4, -0.2 & -0.2, 0.0 & 0.0, 0.2& 0.2, 0.4 & 0.4, 0.6 & 0.6, 0.8 & 0.8, 1.0 & 1.0, 1.2\\
		\hline\hline
		PT & 7.2 & 108.0 & 441.2 & 782.8 & 503.8 & 114.1 &	44.8 & 41.8\\
		FP & 0.1 & 2.2 & 13.2 & 29.7 & 39.5 & 33.5 & 24.9 & 15.9\\
		\hline
		\%FP & 7.1 & 1.9 & 2.9 & 3.7 & 7.3 & 22.9 & 35.7 & 27.7\\
		%\%FP & 4.7 & 1.2 & 2.3 & 3.8 & 6.9 & 20.9 & 38.5 & 25.4\\
		\hline\hline
	\end{tabular}
	\label{tab:3}
\end{center}
\end{table}

\begin{table}
\caption{Best fit variables for PT, FP and \%FP datasets. (
Lin=linear, Poly=polynomial, Exp=exponential, H=horizontal, V=vertical. $\Delta$ is the change of the variable over $50\circ$ longitude or latitude.)
}

    \begin{center}	
    \begin{tabular}{l c c c c c c c c c c c c r}
		%\hline\hline
		&  & V strip & & & & & H strip & & & \\
		%\hline
		Planet radii & Dataset & Best & $\Delta$BIC & $\Delta$BIC & $\Delta$BIC & $\Delta$ & Best & $\Delta$BIC & $\Delta$BIC & $\Delta$BIC & $\Delta$ & \\
		log($R/$R$_{\oplus})$ & & Fit & Poly & Exp & Lin & (50$^{\circ}$) & Fit & Poly & Exp & Lin & (50$^{\circ})$ & \\
		\hline\hline
		-0.4, -0.2 & PT & Exp & 0.60 & - & 0.01 & -0.24 & Exp &  0.74 &  - & 0.14 & -0.25 \\
		 &  FP & Exp & 0.66 & - & 0.01 & -0.01 & Exp & 0.58 & - & 4.11 & -19.55 \\
		& \%FP  & Exp  &  0.75  &  -  & 0.27 & -0.56 & Poly & - & 0.58 & 4.70 &  -6.53  \\
		\hline
		-0.2, 0.0 & PT & Lin & 0.95 & 1.43 & - & +0.31 & Lin & 4.65  &  0.02 &  - & +0.91\\
		&  FP & Lin &  2.36  &  0.43  & 0.08 & +0.31 & Exp &  4.66 &  - & 1.26  & -1.03\\
		& \%FP  & Poly  &  -  & 0.51 & 0.43 & +2.83 & Exp & 2.74 & - & 12.55 & -12.50 & \\
		\hline
		0.0, 0.2 & PT & Exp & 1.02 & - & 0.08 & -5.74 & Lin & 5.11 & 0.01 & - & -2.92 \\
		&  FP & Exp & 0.93 & - & 0.08 & +0.94 & Exp & 4.14 & - & 15.09 & -8.48\\
		& \%FP & Exp & 1.02 & - & 0.08 & +1.62 & Exp & 17.21 & - & 4.89 & -17.14 \\
		\hline
		0.2, 0.4 & PT & Lin & 1.00 & 0.00 & - & -7.71 & Exp & 5.16 & - & 0.36 & -41.57\\
		&  FP & Exp & 0.26 & - & 1.31 & +2.19 & Exp & 11.04 & - & 36.89 & -27.81\\
		&  \%FP & Lin & 0.99 & 0.24 & - & +2.92 & Exp & 3.45 & - & 18.74 & -21.36 \\
		\hline
		0.4, 0.6 & PT & Exp & 0.97 & - & 0.15 & -6.82 & Exp & 5.04 & - & 0.39 & -32.78\\
		& FP  &  Exp &  0.98  & - & 0.02 & +2.59 & Exp & 14.92 & - & 52.51 & -27.10\\
		&  \%FP & Exp  &  1.04  & - & 0.10 & +3.75 & Exp & 5.35 & - & 31.69 & -30.36 \\
		\hline
		0.6, 0.8 & PT & Exp & 0.95 & - & 0.04 & -1.36 & Lin & 4.91 & 0.25 & - & -6.41\\
		& FP  & Exp  &  0.89  & - & 0.02 & +1.25 & Exp & 34.95 & - & 68.65 & -26.47 \\
		& \%FP  & Lin  & 0.06 & 0.02 & - & +7.60 & Exp & 29.78 & - & 46.54 & -67.94 \\
		\hline
		0.8, 1.0 & PT & Exp & 1.12 & - & 0.00 & -0.18 & Exp & 5.28 & - & 0.06 & -3.85\\
		& FP  & Exp & 1.00 & - & 0.01 & +0.03 & Exp & 13.72 & - & 49.08 & -17.32 \\
		& \%FP & Exp & 0.90 & - & 0.02 & +0.74 & Exp & 0.03 & - & 27.29 & -74.73 \\
		\hline
		1.0, 1.2 & PT & Exp & 0.36 & - & 0.02 & -0.22 & Exp & 5.22 & - & 0.02 & -2.74\\
		& FP & Lin & 1.01 & 0.03 & - & +0.54 & Exp & 11.02 & - & 23.53 &  -10.33 \\
		& \%FP & Exp & 0.29 & - & 0.30 & +8.36 & Exp & 17.60 & - & 31.64 & -65.78\\
		\hline
	\end{tabular}
	\label{tab:4}
	\end{center}
\end{table}
%\twocolumn

\begin{table}
\caption{Percent of false positive detections (\%FP), by Galactic latitude ($b$), in `horizontal'\\strips in the default Southern PLATO field (LOPS0)}
    \begin{center}
    \begin{tabular}{l c c c c c c c r}
        %\hline\hline
        \fontsize{10pt}{0}
		& & & & Planet radius & & &\\
		& & & & $\log(R/$R$_{\odot})$ & & &\\
		 $b^{\circ}$ & -0.4, -0.2 & -0.2, 0.0 & 0.0, 0.2& 0.2, 0.4 & 0.4, 0.6 & 0.6, 0.8 & 0.8, 1.0 & 1.0, 1.2\\
		\hline\hline
		-5.5, -9.5  & 27.07 & 11.48 & 12.07 & 14.92 & 23.02 & 57.31 & 68.64 & 59.83\\
		-9.5, -13.5 & 0 & 5.65 & 9.26 & 8.19 & 15.18 & 36.93 & 54.35 & 46.10\\
		-13.5, -17.5 & 6.67 & 9.92 & 3.78 & 5.27 & 9.22 & 23.16 & 41.62 & 36.32\\
		-17.5, -21.5 & 6.67 & 3.13 & 4.46 & 3.14 & 6.77 & 15.67 & 40.99 & 22.15\\
		-21.5, -25.5 & 0 & 1.14 & 2.79 & 4.21 & 5.49 & 15.40 & 32.51 & 18.52\\
		-25.5, -29.5 & 0 & 2.00 & 1.49 & 2.52 & 3.34 & 8.85 & 28.28 & 19.99\\
		-29.5, -33.5 & 0 & 0.93 & 1.17 & 1.56 & 2.80 & 11.94 & 17.42 & 40.26\\	
		-33.5, -37.5 & 0 & 0 & 1.25 & 0.88 & 1.89 & 8.16 & 12.49 & 7.63\\
		-37.5, -41.5 & 6.79 & 0.32 & 0.87 & 1.40 & 2.03 & 8.59 & 29.28 & 18.30\\
		-41.6, -45.5 & 0.21 & 0.22 & 1.05 & 1.06 & 1.70 & 12.91 & 17.42 & 22.26\\
		-45.5, -49.5 & 0 & 0.40 & 0.32 & 0.65 & 1.59 & 3.30 & 24.41 & 21.97\\
		-49.5, -53.5 & 0 & 0.24 & 0.54 & 0.78 & 1.11 & 2.97 & 3.35 & 1.83\\
		\hline\hline
	\end{tabular}
	\label{tab:5}
\end{center}
\end{table}
\twocolumn

The remaining best fits were predominantly exponential with six instances where the linear functions were preferred. However, we note that the preference of one distribution over the other two is very weak in all cases. 

To highlight the magnitude of the change of \%FP with Galactic longitude $l$, we calculate the change in PTs, FPs and \%FPs, per $50^{\circ}$ of $l$ using the best-fit equations.

A graphical representation of the results for the vertical strip analysis for the three radius bins covering the radius range $-0.2<\log(R/$R$_\oplus)\leqslant0.4$, are shown in the left hand panels of Figure \ref{fig:plots}. 

Considering the trend across the $50^{\circ}$ Galactic longitude step calculated above, we find a modest \textbf{negative} correlation across all the radii bins between the number of PTs and increasing $l$ amounting to no more than $\sim$ 3\% across the $50^{\circ}$ step. For the FPs we find a modest \textbf{positive} correlation between the number of FPs and increasing $l$, again amounting to no more than $\sim$ 3\% across the $50^{\circ}$ step. 
To test this correlation, we calculated the best-fit flat-line (i.e. no correlation between $l$ and the \%FP), for the planet radius bins $-0.2<\log(R/$R$_{\odot})<0.8$. Table \ref{tab:flat} shows that for the radius range analysed, the exponential best-fit function is only marginally preferred over the flat-line  best-fit. Analysing the best-fit and the flat-line best-fit  functions using the Bayesian information criteria ($\Delta$BIC), we find no evidence for the exponential function to be preferred over a flat \%FP value for the radius bins $0.0<\log(R/$R$_{\odot})<0.6$ and only a slight preference for the exponential best-fit function  over the flat-line best fit for the other two radius bins.

We conclude that while there appears to be a weak positive correlation between the Galactic longitude $l$ and the \%FP, and we would expect such a correlation as we move closer to the Galactic centre, this is not strongly favoured statistically over no trend.

\begin{table}
    \caption{Best fit functions for \%FP for vertical strips compared to no trend (\%FP constant). $\sigma_\textrm{res}$ is the mean of the residuals.}
    \begin{center}
    \begin{tabular}{l c c c c c c c r}
		%\hline\hline
		Planet radius\\ $\log(R/$R$_{\odot})$ & -0.2, 0.0 & 0.0, 0.2& 0.2, 0.4 & 0.4, 0.6 & 0.6, 0.8 & \\
		\hline\hline
		Best fit \%FP  & Exp & Exp & Exp & Exp & Exp \\
		$\sigma_{\textrm{res}}$ & 0.35 & 0.20 & 0.26 & 0.33 & 1.06\\
		\hline
		\%FP = const & 2.3  & 3.4 & 4.3 & 7.7 & 22.8 \\
		$\sigma_\textrm{res}$ & 0.44 & 0.26 & 0.38 & 0.53 & 1.31\\
		\hline
		$\Delta$BIC & 5.0 & 1.9 & 0.8 & 0.0 & 2.9\\
		\hline\hline
	\end{tabular}
	\label{tab:flat}
\end{center}
\end{table}

  \subsection{LOPS0 - analysis of strips of constant latitude}\label{sclat}
 We now turn to the horizontal strips which allow us to probe the latitude dependence of the false positive rate as we move up the LOPS0 field. 
 
 For the PTs, FPs and \%FPs there is again only one radius bin where a polynomial fit is preferred. Almost exclusively, the exponential distribution is preferred, and in most cases the preference for this distribution over the others is strong, especially for the \%FP datasets. 
  As expected, we see a significant reduction in the \%FP with increasing $|b|$, as the stellar density reduces and the target background and foreground fields become less dense.
We once again calculate the change in \%FP per $50^{\circ}$, this time for the Galactic latitude $b$ using the best-fit equations. 

The magnitude of the change in \%FP is significant, with the decrease with $|b|$ being approximately 10 times the increase in \%FP for the corresponding radius bin. Of particular note is the significant increase in the \%FP in each radius bin in the region $|b|\leqslant13.5^{\circ}$ (refer to Table \ref{tab:5}). 
A graphical representation of the results for the horizontal strip analysis for the three radius bins covering the radius range $-0.2<\log(R/$R$_\oplus)\leqslant0.4$, is shown in the right hand panels of Figure \ref{fig:plots}. %(Note again the changing \%FP y-axis scales).
 
 While the overall \%FP in the LOPS0 field is relatively low, the increase in the \%FP near the Galactic plane is significant. For the three bins of most interest, the average \%FP is approximately 13\% in the region $5.5<|b|\leqslant 9.5^{\circ}$. 
 
While it is tempting to compare the dependence of the PLATO \%FP on Galactic longitude and latitude to what is seen for false positives in the Kepler field, a meaningful comparison is difficult without considerable analysis. This is mainly because the instrument characteristics are very different between Kepler and PLATO; in particular, Kepler's angular resolution is about four times better, and so most eclipsing binaries that give rise to FPs for PLATO are irrelevant for Kepler. 

While such a detailed analysis is beyond the scope of this paper, we have carried out a basic analysis to calculate the approximate Kepler \%FP rate by Galactic latitude. We extracted the Kepler confirmed planets and false positives from the NASA Exoplanet Science Institute (NExScI)   
 catalogue, limiting both datasets to a maximum Kepler magnitude $K_p<13$, a maximum planet radius mimic of $\log(R/$R$_{\odot})<1.2$ and combining all periods. For the false positives we consider only those identified as being a result of eclipsing binaries to match our synthetic analysis. We binned the \%FP into three bins, $9.5<|b|\leqslant 13.5^{\circ}$, $13.5<|b|\leqslant 17.5^{\circ}$ and $17.5<|b|\leqslant 21.5^{\circ}$, to match our synthetic data bins. For the synthetic \%FP we have simply averaged the \%FP for the relevant latitude range shown in Table \ref{tab:5}.

\begin{table}
    \caption{Comparison of the \%FPs in the synthetic PLATO field versus those observed by Kepler.}
    \begin{center}
    \begin{tabular}{l c c r}
		%\hline\hline
		Latitude range & PLATO SPF0 & Kepler\\
		\hline\hline
		9.5 : 13.5 & 22\% & 18\%\\
		\hline
		13.5 : 17.5 & 17\% & 14\%\\
		\hline
		17.5 : 21.5 & 13\% & 7\%\\
		\hline\hline
	\end{tabular}
	\label{tab:kvsspf0}
\end{center}
\end{table}

 As can be seen from Table \ref{tab:kvsspf0}, we find that the trend of the \%FP with Galactic latitude in our synthetic SPF0 field is qualitatively consistent with that seen in the Kepler field. As expected from the difference in angular resolution, the \%FP seen in our synthetic PLATO field is higher than for Kepler. We reiterate that our analysis is only approximate and a more rigorous analysis would be required to confirm the match.

 \section{Extrapolation to other field centre locations within the allowed LOPS region}\label{extrap}
 Given the random nature of the orbital alignments, the more target stars that are observed the higher the expected number of planet detections. As a result, the prime reason for investigating the effect of different field centre coordinates on \%FP is to determine at what point the increased number of target stars observed by moving to a more crowded field, is negated by the increase in false positives resulting from more densely populated background and foreground fields. 
 
 Since our results are obtained from a synthetic population based on statistical distributions, and are subject to numerous simplifications and parameterizations of the input physics, we recognize that the simulated absolute number of PT and FP detections are subject to uncertainties and may well differ from the absolute number of future observed detections. However, by using a ratio of FP/(FP + PT), the effect of simplifications in the detector modelling and assumptions in the input physics is applied equally to the numerator and denominator and hence effectively cancel out.

 To quantify the effect on the \%FP of moving the field centre location within the allowed region, we use the best-fit \%FP distributions from the vertical strips to model the effect on the \%FP of changing the Galactic longitude of the field centre location while keeping the latitude constant. Similarly we use the best-fit \%FP distributions from the horizontal strips to model the effect on changing the Galactic latitude of the field centre while keeping the Galactic longitude constant. For simplicity, where a region of LOPS0 crosses into the northern Galactic hemisphere we model the Galactic latitude using the corresponding southern hemisphere latitude, i.e. the $b = +4^{\circ}$ datasets are modelled using the best-fits for $b = -4^{\circ}$. We then scale the resultant \%FP by the relative combined area of the sub-fields and the synthetic LOPS0. 
 
 Uncertainties for each extrapolated point $\hat{y}(\hat{x})$ were calculated using $\sigma_{\hat{y}}=\sqrt{\frac{\sigma^2}{N}\times \left[1+\left(\frac{   \hat{x}-\overline{x}}{\sigma_x} \right) ^2 \right]}$ \footnote{https://w3.pppl.gov/~hammett/work/1999/stderr.pdf}. While the \cite{bevington2003data} formula this equation is based on is for a linear fit, we believe, given the other uncertainties in our data,  this provides an acceptable estimate for the uncertainties in the extrapolated points. The total uncertainties for each planet radius bin for the new field locations were obtained by combining the uncertainty in the \%FP in SPF0, the uncertainty in the best fit for the \%FP in SPF0 and each uncertainty in the extrapolated points using the sum of the squares.

\begin{figure}
    \centering
    \includegraphics[width=\hsize]{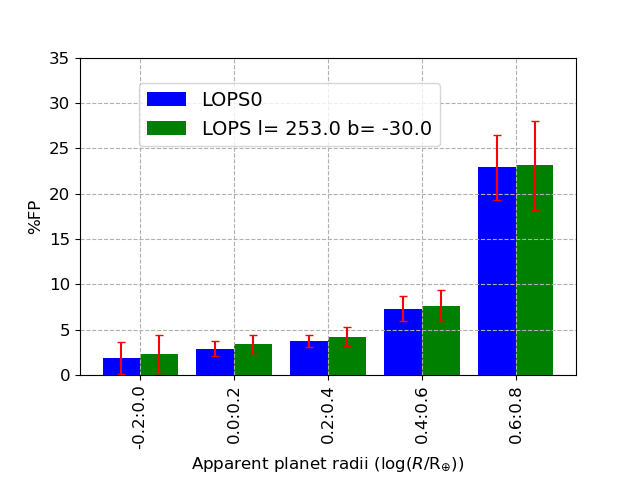}
    \caption{Comparison of the projected false positive percentage (\%FP), by planet radius bin for LOPS0 ($l=253^{\circ}$ $b=-30^{\circ}$). The blue bars show the LOPS0 \%FP using the camera coverage pattern shown in Figure \ref{fig:FOV} and the green bars are the \%FP projected using $\mathbf{vertical}$ strip best-fit equations for the LOPS0 field assuming the entire area is covered with two cameras.}
    \label{fig:253/30}
\end{figure}

\begin{figure}
    \centering
    \includegraphics[width=\hsize]{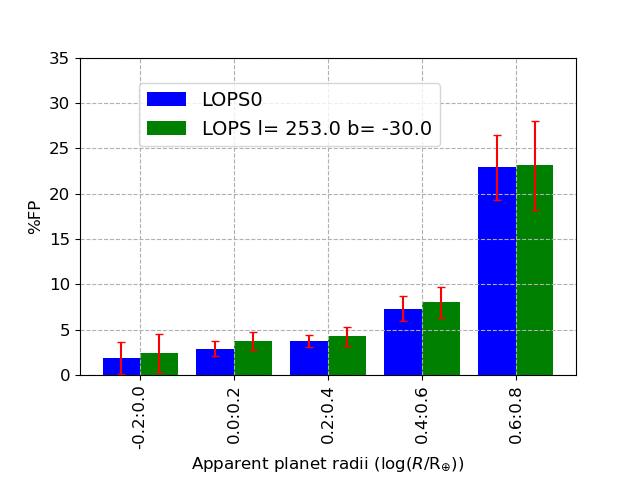}
    \caption{Comparison of the projected false positive percentage (\%FP), by planet radius bin for LOPS0 ($l=253^{\circ}$ $b=-30^{\circ}$). The blue bars show the LOPS0 \%FP using the camera coverage pattern shown in Figure \ref{fig:FOV} and the green bars are the \%FP projected using $\mathbf{horizontal}$ strip best-fit equations for the LOPS0 field assuming the entire area is covered with two cameras.}
    \label{fig:253/30Sqr}
\end{figure}
\begin{figure}
    \centering
    \includegraphics[width=\hsize]{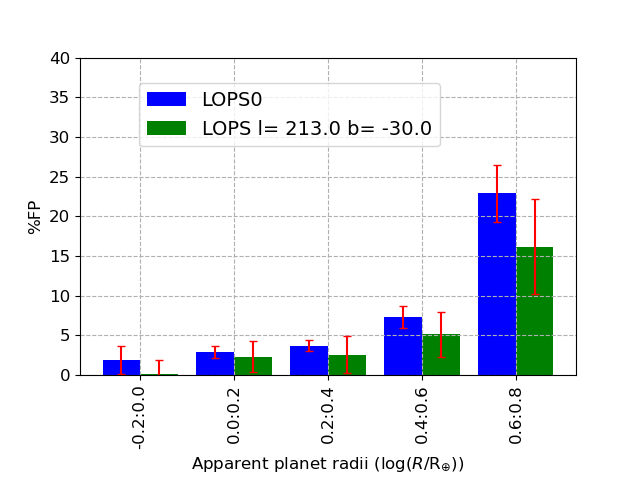}
    \caption{The projected false positive percentage (\%FP), by planet radius bin when centre Galactic longitude only changes. Blue bars show the LOPS0 \%FP and the green bars are those projected \%FP for an LOPS field centre of $l=213^{\circ}$ ($b=-30^{\circ}$)}
    \label{fig:213/30}
\end{figure}
\begin{figure}
    \centering    
    \includegraphics[width=\hsize]{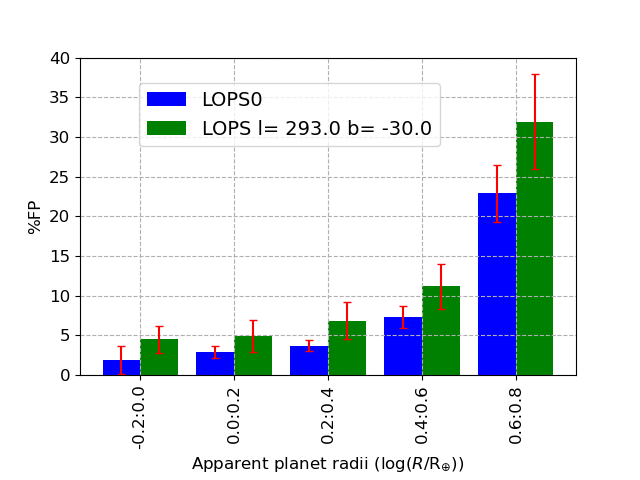}
    \caption{As per Figure \ref{fig:213/30} but with green bars showing the projected \%FP for an LOPS field centre of $l=293^{\circ}$ ($b= -30^{\circ}$)}
    \label{fig:293/30}
\end{figure}

In Figure \ref{fig:253/30} we show the comparison between the predicted \%FP by apparent planet radii mimicked using the 15 LOPS0 simulations and camera overlap pattern shown in the right hand panel of Figure \ref{fig:FOV} (blue bars), and that obtained by combining the vertical strips (green bars), and scaling the area to match LOPS0. In Figure \ref{fig:253/30Sqr} we show the same match this time using the horizontal strips.  The approximation in the strip analysis of each FP, regardless of its location, being observed by two camera groups has little effect on the results, with the \%FP observed in the combined and scaled strips differing by no more than $\approx$1\% when compared to the results obtained using the LOPS0 shape and overlap pattern. Again we have excluded the smallest planet radius bin due to the small number statistics involved and exclude the two largest planet radius bins with $\log(R/$R$_\oplus)>0.8$, as these are well outside the planet radius range of interest to the PLATO mission. 

As expected, the effect on \%FP of changing the Galactic longitude $l$ coordinate of the field centre location is minor with only a relatively modest change in the \%FP observed as the Galactic longitude $l$ field centre location is moved away from, or towards, the Galactic centre by $40^{\circ}$ (Figures \ref{fig:213/30} and \ref{fig:293/30} respectively).

As highlighted in Section \ref{taspf0}, the effect of changing the field centre Galactic latitude $b$ on the \%FP is far greater than changing the field centre coordinate galactic longitude $l$. 

\begin{figure}
    \centering
    \includegraphics[width=\hsize]{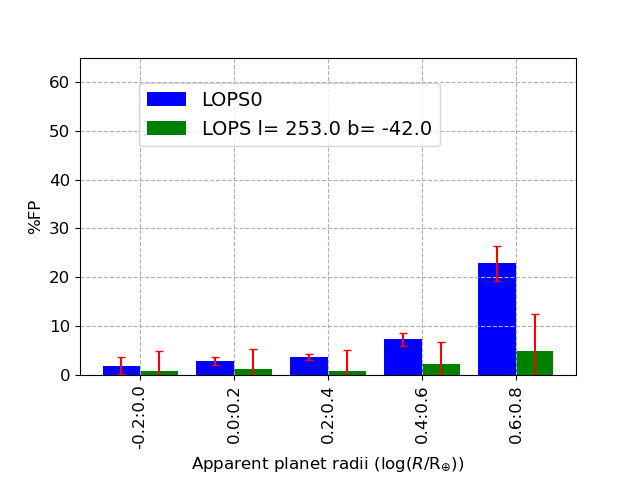}
    \caption{The projected false positive percentage (\%FP), by planet radius bin when centre latitude only changes. Blue bars show the LOPS0 \%FP and the green bars are those projected for an LOPS field centre of $b= -42^{\circ}$ ($l=253^{\circ}$)}
    \label{fig:253/40}
\end{figure}

While we show the effect on \%FP for a field centre location of Galactic latitude $b = -42^{\circ}$ in Figure \ref{fig:253/40}, this is for completeness only as such a location would most likely be rejected as the total number of target stars observed in the two LOP fields is expected to drop well below the minimum requirement of 267,000 \citep{Rauer_2014}. 

\begin{figure}
    \centering
    \includegraphics[width=\hsize]{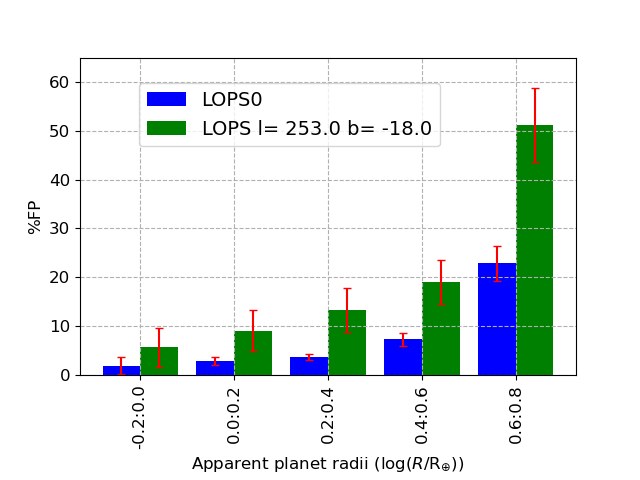}
    \caption{The projected false positive percentage (\%FP), by planet radius bin when centre latitude only changes. Blue bars show the LOPS0 \%FP and the green bars are those projected for an LOPS field centre of $b= -18^{\circ}$ ($l=253^{\circ}$)}
    \label{fig:253/20}
\end{figure}

\begin{figure}
    \centering
    \includegraphics[width=\hsize]{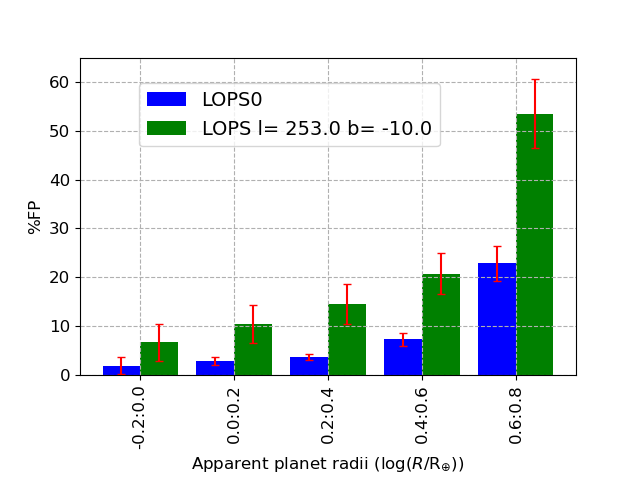}
    \caption{As per Figure \ref{fig:253/20} but with green bars showing the projected \%FP for an LOPS field centre of $b= -10^{\circ}$ ($l=253^{\circ}$).}
    \label{fig:253/10}
\end{figure}

In Figures \ref{fig:253/20} and \ref{fig:253/10} we show the effect of moving the field centre closer to the Galactic plane, galactic longitude $l= 253^{\circ}$ Galactic latitude $b= -18^{\circ}$ and Galactic longitude $l= 253^{\circ}$ Galactic latitude $b= -10^{\circ}$ respectively. As expected, these result in a significant increase in the \%FP. We note that a smooth Galactic density model is not likely to be representative of the actual population at very low latitudes and caution that statistics from  locations of $|b|\leqslant5^{\circ}$ need to be regarded as indicative only. 


While the effect of moving both the Galactic longitude $l$ and Galactic latitude $b$ locations of the field centre could be modelled by averaging the effects of our two best-fit functions, the averaging would introduce more uncertainty and given the small effect on the \%FP with changing galactic longitude $l$, such an exercise was not deemed to be beneficial.

\section{Comparison to LOPN and PICTarget110}\label{LOPSvsLOPN}
We wish to compare our LOPS0 results with LOPN0. To this end we rendered approximately 75\% of LOPN0 covering Galactic coordinates $l=40.5^{\circ}$ to $l=89.5^{\circ}$ and $b=5.5^{\circ}$ to $b=53.5^{\circ}$ which we call LOPN-sub. We create a comparable area in the LOPS covering $l=228.5^{\circ}$ to $l=277.5^{\circ}$ and $b=-5.5^{\circ}$ to $b=-53.5^{\circ}$ which we refer to as LOPS-sub. This enables us to compare and contrast the properties of these sub fields. A complete renditon of LOPN0 is time-consuming and beyond the scope of this study.

\subsection{Stellar Populations}
We first compare the P5 target star population of PICTarget110 to the corresponding regions in our LOPS-sub and LOPN-sub fields. To remove the effects of the PLATO detector geometry on the field shape we create sub fields from PICTarget110 and consider only the central region of the PICTarget110 northern and southern fields. We refer to these sub-fields as PIC-LOPS-mini and PIC-LOPN-mini. The synthetic fields matching the PIC-mini fields comprise approximately 30\% of the LOPs and cover $l=243^{\circ}$ to $l=263^{\circ}$ and $b=-9.5^{\circ}$ to $b=-53.5^{\circ}$ (LOPS-mini) and $l=55^{\circ}$ to $l=75^{\circ}$ and $b=9.5^{\circ}$ to $b=53.5^{\circ}$ in (LOPN-mini). Again for simplicity, we assume the entire area is observed by two camera groups.

\begin{table}
	\centering
	\caption{P5 target stars and unresolved binaries in PIC-LOPS-mini, PIC-LOPN-mini, LOPS-mini and LOPN-mini}
	\label{tab:targcomp}
	\begin{tabular}{lccr} % four columns, alignment for each
		\hline
		Field & Single stars & Unresolved binaries & Total\\
		\hline
		PIC-LOPN-mini & 51563 & - & 51563\\
		LOPN-mini & 23559 & 25079 & 48638\\
		PIC-LOPS-mini & 47486 & - & 47486\\
		LOPS-mini & 24947 & 26113 & 51060\\
		\hline
	\end{tabular}
\end{table}

As in Section \ref{spfaS1}, when we consider the unresolved binaries in our synthetic fields as target stars, we find a very good correlation between the target star numbers in PIC-LOPS-mini and LOPS-mini and between those in PIC-LOPN-mini and LOPN-mini. Both show variations of less than 10 percent (see Table \ref{tab:targcomp}). While our synthetic fields show more target stars in LOPS-mini than LOPN-mini, which is the opposite of that seen in PIC-mini datasets, we again find less than a 10 percent variation which we believe is an acceptable variation given the approximations in the Galactic structure and input physics used in our simulations.

\begin{figure}
    \centering
    \includegraphics[width=\hsize]{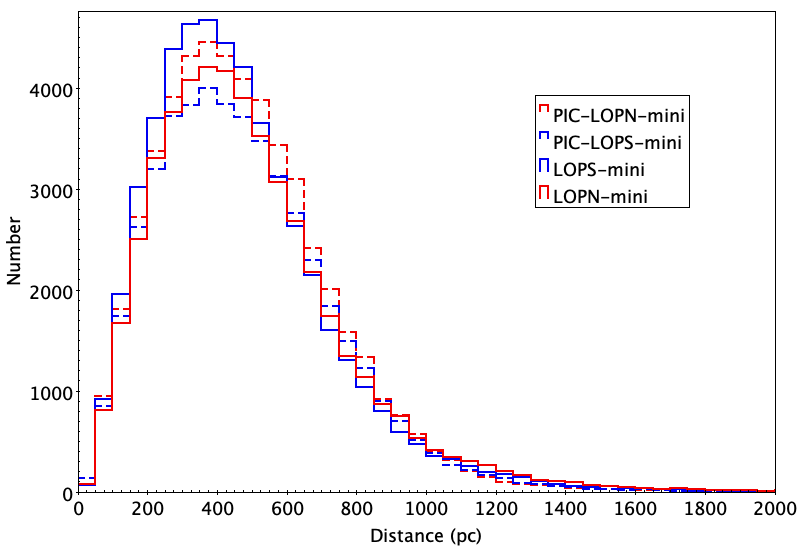}
    \caption{Comparison of target star distances in the smaller long duration observational field subsets - LOPS-mini solid blue line, LOPN-mini solid red line, PIC-LOPS-mini dashed blue line and PIC-LOPN-mini dashed red line.}
\label{fig:SNDist}
\end{figure}

As well as an acceptable correlation between the target star numbers in the PIC and synthetic mini fields we compare the distance distribution of the target stars in PIC-mini and synthetic mini fields. Again we find a good correlation with both the peak target star distance and overall distribution shape a good match. (see Figure \ref{fig:SNDist}).

\subsection{PTs and FPs in the LOPS-sub and LOPN-sub fields}
Returning to the LOPS-sub and LOPN-sub fields, we find a total of 1,749 PTs in LOPN-sub compared to 1,917 in LOPS-sub. Both PT and FP numbers were obtained by averaging results from five simulations with the error bars obtained from the standard deviations.

\begin{figure}
    \centering
    \includegraphics[width=\hsize]{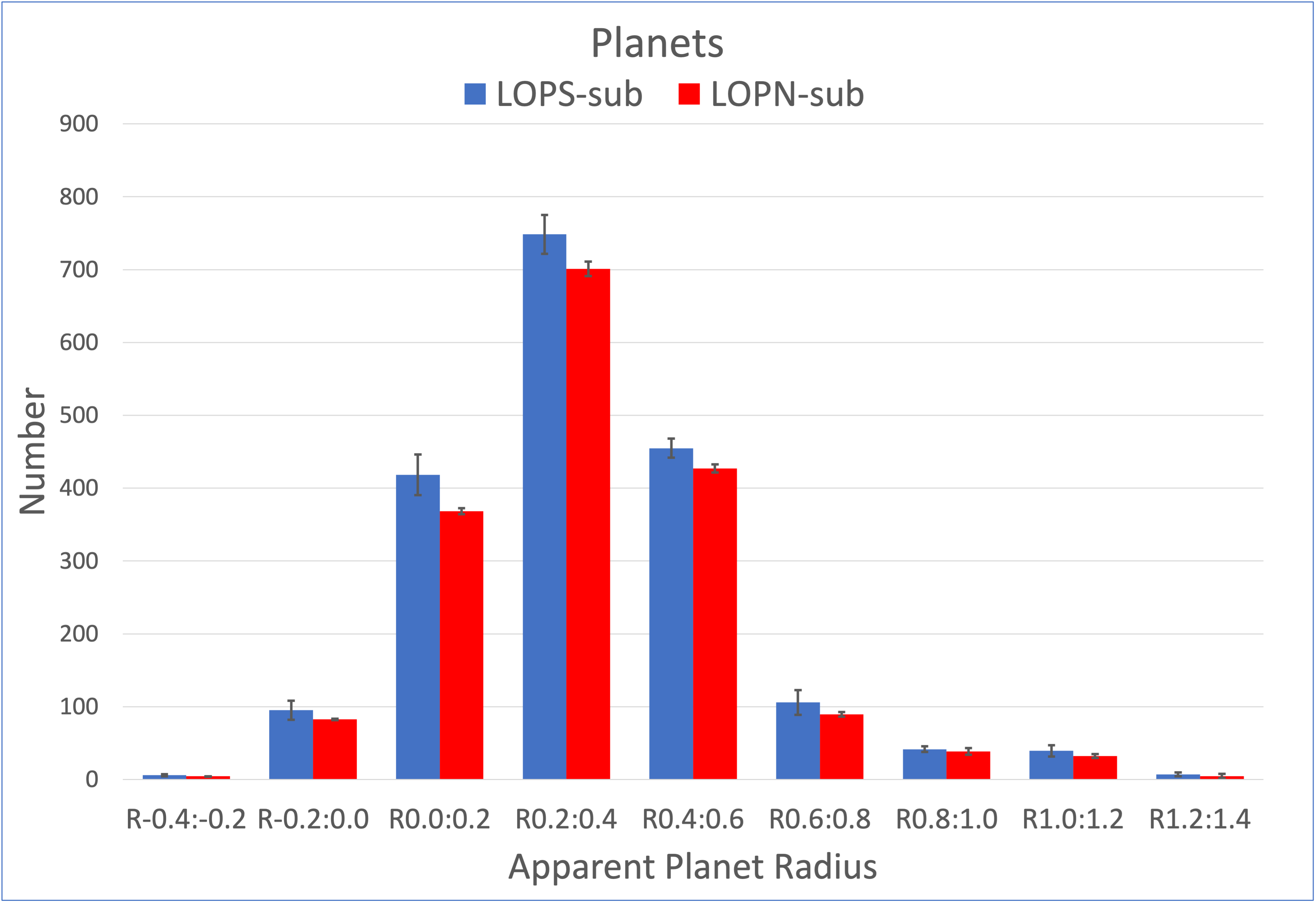}
    \caption{Comparison of planetary transit numbers (PTs) by radii detected in the long duration observational field south sub field (LOPS-sub) and the long duration observational field north sub field (LOPN-sub)}
    \label{fig:LOPSLOPNPLANET}
\end{figure}

As shown in Figure \ref{fig:LOPSLOPNPLANET} despite the slight variation in PT numbers we see a very similar apparent planetary radius distribution in LOPS-sub and LOPN-sub. 

\begin{figure}
    \centering
    \includegraphics[width=\hsize]{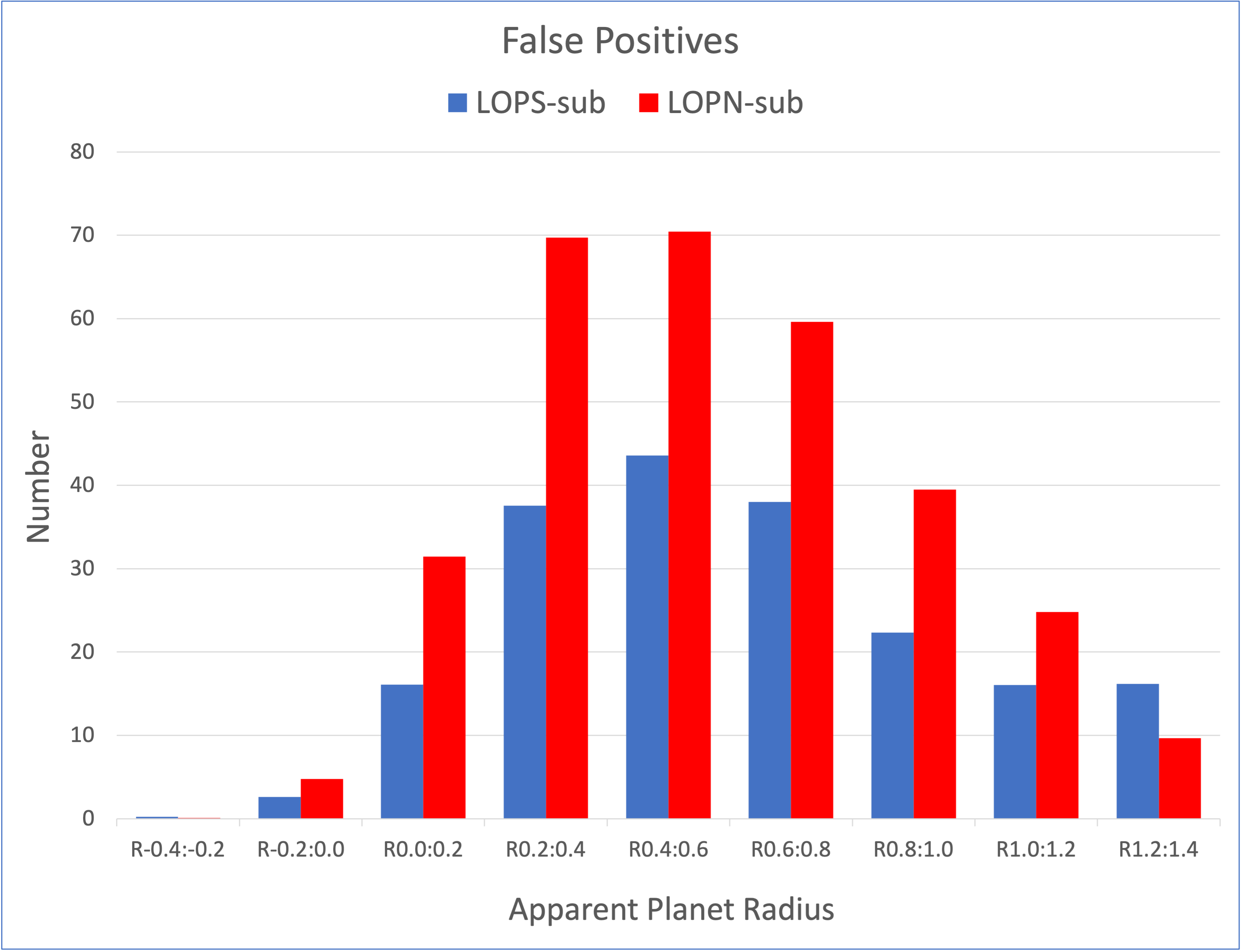}
    \caption{Comparison of false postive planetary transit numbers (FPs) detected in the long duration observational field south sub field (LOPS-sub) and the long duration observational field north sub field (LOPN-sub)}
    \label{fig:FPnumbers}
\end{figure}

While the PT numbers in LOPS-sub and LOPN-sub are very similar the same cannot be said for the FP numbers. We find considerably more FPs in the LOPN-sub field as shown in Figure \ref{fig:FPnumbers}.

\begin{figure}
    \centering
    \includegraphics[width=\hsize]{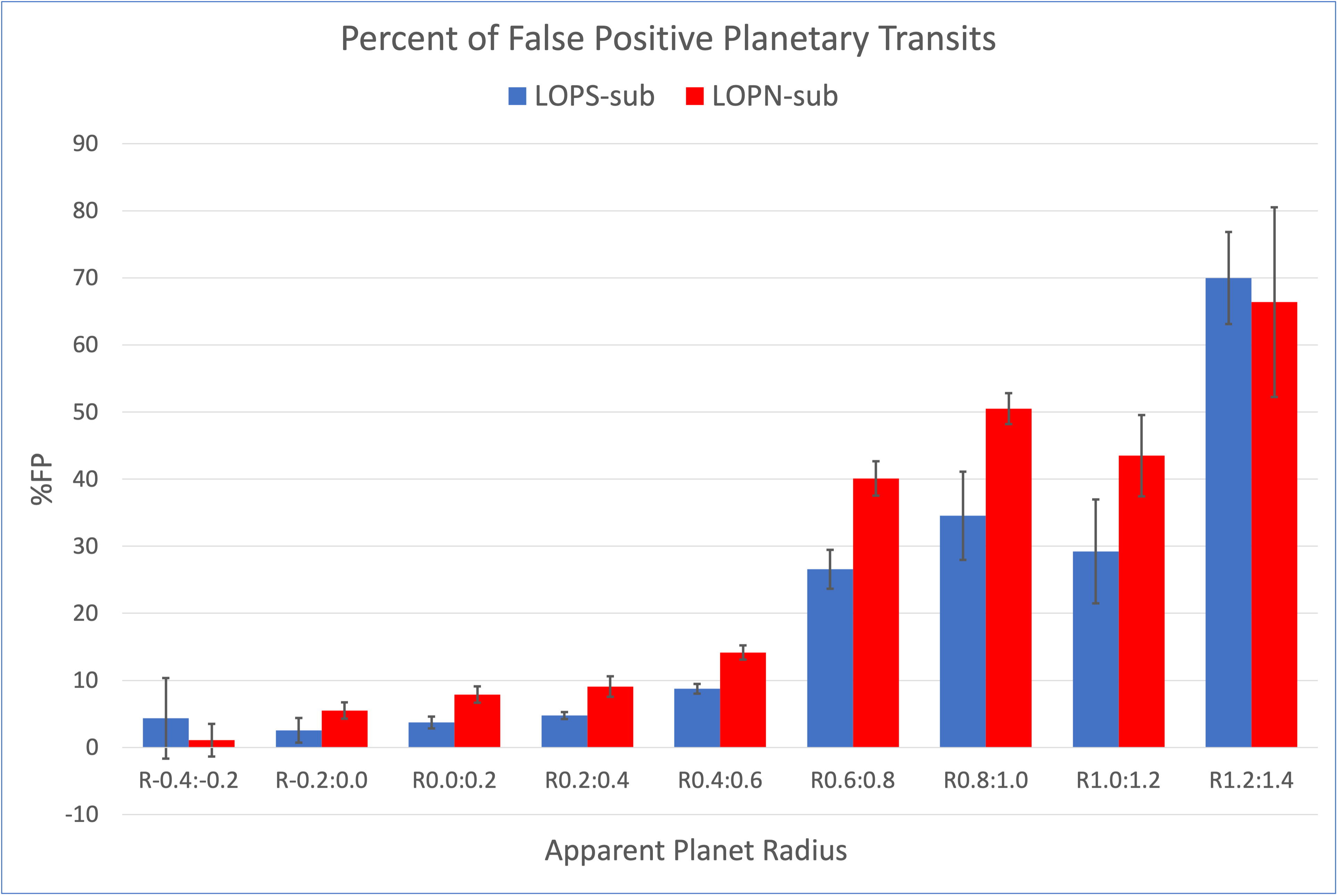}
    \caption{Comparison of false postive planetary transits, in percent, detected in the long duration observational field south sub field (LOPS-sub) and the long duration observational field north sub field (LOPN-sub)}
    \label{fig:FPsub}
\end{figure}

The combined effect of slightly lower PTs and higher FPs in LOPN-sub results in a considerable higher \%FP in LOPN-sub than in LOPS-sub as shown in Figure \ref{fig:FPsub}. 

\section{Discussion and Conclusions}\label{dc}
In this work we have used the \textsc{BiSEPS} synthetic stellar models and population synthesis code to create a complete synthetic single and binary stellar population for LOPS0. The synthetic population was calibrated to reproduce the Kepler field content in $\log(g)$ vs $\log(T_{eff})$ space, as well as the mass ratio and period distribution of Kepler-identified eclipsing binaries. We have verified that the number of synthetic single P5 target stars in the resulting synthetic LOPS0 is comparable to that in the PLATO input catalogue PICTarget110 when an  estimate of the unresolvable binary population is added to the synthetic single star population and that the target star distribution by Galactic latitude and longitude is a good match to PICTarget110. Using the same calibration settings, we repeated the analysis for a subfield of the LOPN and again find a good correlation with PICTarget110 both in target star numbers and distance distribution. We reiterate that our FPs are those predicted from blended eclipsing binaries only and we do not consider other potential false positive signals such as from stellar activity.

We seeded each single P5 target star in the synthetic LOPS0 super-pixels and in LOPN-sub super-pixels with a single planet whose number, radius and period distributions have been calibrated to reproduce the Kepler planet sample from the synthetic Kepler field.  

We have analysed the resulting detectable PTs and FPs  from blended eclipsing binaries and, using simplified specifications for the PLATO instrument, determined the false positive rate by apparent planetary radius when each signal is blended with four PLATO pixels.

\subsection{LOPS}
In the radius range of most interest to PLATO,   $-0.2<\log(R/$R$_{\oplus})\leqslant0.4$, we predict an average a false positive rate of approximately $3\%$ from blended eclipsing binary systems in the LOPS0 field. However, we note a significant increase in the false positive rate in regions of the LOPS0 where the absolute value of the Galactic latitude $|b| \leqslant 9.5^{\circ}$, and we show that most planetary radius bins in the synthetic LOPS0 would see a false positive rate of at least 20\% for regions where Galactic latitude $|b| \leqslant 5.5^{\circ}$. 

We split our synthetic LOPS0 into a number of equal area vertical and horizontal sub-fields to analyse the effect of varying the field centre Galactic longitude $l$ or Galactic latitude $b$ coordinates respectively on the PTs, FPs and false positive rate. 

The vertical sub-field analysis indicates a modest increase in the false positive rate with field centre longitude $l$. We obtain the best fit functions to model these trends and use these equations to estimate the false positive rate expected for alternative longitude $l$ field centre locations while keeping the latitude $b$ field centre location constant. We conclude that the Galactic longitude of the field centre location within the LOPS allowed region does not significantly affect the false positive rate from eclipsing blended binaries and as a result, conclude that, from the perspective of astrophysical false positives, the Galactic longitude choice is not a significant factor in the selection of the LOPS field centre.

Repeating a similar trend analysis on the horizontal strips and using the best-fit functions to model the false positive rate for alternative Galactic latitude $b$ field centre locations while keeping the Galactic longitude $l$ field centre location constant, we find a more significant increase in the false positive rate with decreasing field centre latitude $|b|$. While this indicates a lower false positive rate will be obtained with higher values of field centre latitude $|b|$, the number of target stars rapidly drops below the mission threshold of 267,000 assuming both the LOPS and the LOPN adopted higher $b$ field centre locations. As a result, latitude pointings of $|b| \geqslant 30^{\circ}$ will most likely not be considered.

For field centre latitude pointings of $|b| \leqslant 30^{\circ}$, our research suggests that while the false positive rate increases significantly, the overall rates are still likely to be less than $\sim$15\% for the planetary radius bins in the range $-0.2<\log(R/$R$_{\oplus})\leqslant0.4$ even for a field centre location of $b = -10$.

This suggests that moving the field centre location to regions where Galactic latitude 
$|b| < 30^{\circ}$, would be advantageous for the planet radius range of most interest, namely $-0.2<\log(R/$R$_{\oplus})\leqslant0.4$.

The effect of field centre latitude pointings of Galactic latitude $|b| < 30^{\circ}$ on larger planets is much more significant and we estimate that over 80\% of $\mathbf{additional}$ transit detections in the region of $|b| \leqslant 5.5^{\circ}$ in the radius range $\log(R/$R$_{\oplus})>0.6$ would be false positives. 

We highlight that splitting the synthetic LOPS0 into sub-fields has resulted in some radius bins having very low counts for FPs, and this has resulted in increased uncertainties in the false positive rate results especially for the smallest planetary radius bin  $-0.4<\log(R/$R$_{\oplus})\leqslant-0.2$. As a result we generally exclude this bin from our discussion but show the strip analysis results in Table \ref{tab:4} for completeness. 

Our intrinsic planet distribution has been calibrated to match the Kepler confirmed planets using an approximation of the Kepler instrument properties. As a result, we believe our planet detection rates for PLATO using an approximation of the PLATO instrument properties provides a good `order of magnitude' estimate for the expected planet detections. In the case of false positives from eclipsing binary systems, we constructed a \textsc{BiSEPS} synthetic stellar population tailored to reproduce the Kepler target list in $\log T_{eff}$ and $\log(g)$ space, used an approximation of the Kepler instrument properties, and then modified the resultant eclipsing binary period distribution to once again match the observed Kepler eclipsing binary data. While we believe both our resulting planet detections and false positive rates provide reasonable approximations for each dataset, our focus is on differential analysis of the \%FP rate with varying Galactic latitude and longitude both within the suggested PLATO field, and then for varied locations of the field within the allowed regions. Therefore we have focused our analysis on the differential change in the false positive rates (\%FP), which we believe is a more robust statistic. This is because using a ratio of FP/(FP + PT), the effect of simplifications in the detector modelling and assumptions in the input physics is applied equally to the numerator and denominator and hence effectively cancel out. As a result, we caution against any inferences made using our PT and FP numbers alone. 

Our method of calculating the transit and the eclipse depths by blending them with the flux from the super-pixel containing the target star is only an approximation of the method suggested in \cite{Marchiori:2019aa}. Their research suggests that both the number and location of the pixels used in the mask will depend on the point spread function, of the target star. Such an analysis is beyond the scope of this paper.

We have not attempted to model the detector in detail, rather, using a standard CCD equation and adopting realistic approximate noise values and throughputs, we have calculated an estimate for what might be expected which we deem broadly representative of the detector characteristics. 

As shown in the example of changing the detection threshold in Section \ref{rspf0}, while the FPs show a reduction of $\approx97\%$ in the planet bin $0.0<\log(R/$R$_{\oplus})\leqslant0.2$, for example, the reduction in PTs in the same bin is similar (a reduction of $\approx98\%$), resulting in a similar ratio. (Note that we are \textbf{not} suggesting the PLATO data processing pipeline should use a one-sigma detection threshold. Rather, that in the region of planet radii of most interest to the PLATO mission ($-0.2<\log(R/$R$_{\oplus})\leqslant0.4$), we find that the detection threshold has little effect on the overall value of the \%FP  calculated using our simulations.)

The main goal of our study is to explore the differential change of the false positive rate (\%FP) with varying Galactic latitude and longitude, both within the suggested PLATO field and then for varied locations of the field within the allowed regions. The differential change of \%FP we observe in our simulations is a more robust prediction than the magnitude of the false positive rate, while \%FP in turn is a more robust result than the simulated absolute numbers of false positives and transiting planets in our model. 

\subsection{LOPN}
While a full comparison of the LOPN field similar to that carried out for LOPS was beyond the scope of this study, the analysis of a significant portion of the LOPN sub-field suggests PLATO can expect a considerably higher false positive rate in LOPN compared to the LOPS. This increase is a result of two factors, firstly we find slightly fewer target stars in the LOPN and secondly we find more FPs in the LOPN. 
While the lower number of target stars in LOPN is at odds with PICtarget110, the variation is not significant and is most likely an artefact from our simplified double disc Galactic structure. Of more significance is the greatly increased number of FPs in LOPN. While this is also driven by our synthetic Galactic structure model, the background in LOPN compared to the LOPS is observably more dense than the LOPS due to the LOPN being oriented more toward the Galactic centre. This is supported by the PICcontaminant110 dataset which shows approximately 36\% more contaminants in the LOPN than the LOPS. \citep{Montalto2021}.

\subsection{Conclusions}
Our research indicates that the \%FP from blended eclipsing binaries using the current proposed LOPS centre location of $b=-30^{\circ}$ and $l=253^{\circ}$, is relatively low with an average of approximately 3\% in the radius range of most interest namely $-0.2<\log(R/$R$_{\oplus})\leqslant0.4$. However, our research suggests that moving the field centre location closer to the Galactic centre to include targets with $|b|\leqslant5.5^{\circ}$ will result in a significant increase in the \%FPs especially in the radius range $\log(R/$R$_{\oplus})>0.6$ where we expect that on average more than 80\% of additional detections will be FPs.  For the LOPN, moving the field centre location closer to the Galactic plane will result in virtually all additional planet detections in the radius range $\log(R/$R$_{\oplus})>0.6$ being false positives.

Countering this increase, we find that only approximately 2.6\% of the identified fully eclipsing FPs have periods between 180 and 1,000 days, resulting in the vast majority of the FPs identified being discounted as planetary transits of Earth-like planets around Sun-type stars due to their short periods. 
If the focus is purely on Earth-like planets orbiting Sun-type stars, moving the LOPS field centre closer to the Galactic plane should result in more target stars and more planet detections but the compromise will be significantly more false positives for shorter period planet transits and larger radius planets.

The difference in false positive rates for LOPN and LOPS suggests that, from an astrophysical false positive perspective, the LOPS would be a more productive field and consideration should be given to dedicating more of the initial four-year observational window to the southern field. 

\section*{Acknowledgements}
The authors would like to thank the anonymous referee for their thorough analysis and insightful suggestions which enabled us to significantly improve the paper. 

This work presents results from the European Space Agency (ESA) space mission PLATO. The PLATO payload, the PLATO Ground Segment and PLATO data processing are joint developments of ESA and the PLATO Mission Consortium (PMC). Funding for the PMC is provided at national levels, in particular by countries participating in the PLATO Multilateral Agreement (Austria, Belgium, Czech Republic, Denmark, France, Germany, Italy, Netherlands, Portugal, Spain, Sweden, Switzerland, Norway, and United Kingdom) and institutions from Brazil. Members of the PLATO Consortium can be found at https://platomission.com/. The ESA PLATO mission website is https://www.cosmos.esa.int/plato. We thank the teams working for PLATO for all their work.\\
\\
UK acknowledges support by STFC grant  ST/T000295/1.\\ 
\\
This research was supported by UKSA grant ST/R003211/1 (Open University element of PLATO UK - Support for the Development Phase)\\ 

JCB acknowledges the support provided by the University of Auckland and funding from the Royal Society Te Ap$\bar{\textrm{a}}$rangi of New Zealand Marsden Grant Scheme.

\section*{Data availability}
Data available on request from the authors.
%\end{acknowledgements}

%The last numbered section should briefly summarise what has been done, and describe
%the final conclusions which the authors draw from their work.

%%%%%%%%%%%%%%%%%%%% REFERENCES %%%%%%%%%%%%%%%%%%

\bibliographystyle{mnras}
\bibliography{PLATO.bib} 

% Don't change these lines
\bsp	% typesetting comment
\label{lastpage}
\end{document}